\documentclass[journal]{IEEEtran}
\IEEEoverridecommandlockouts
\makeatletter
\def\endthebibliography{%
  \def\@noitemerr{\@latex@warning{Empty `thebibliography' environment}}%
  \endlist
}
\makeatother

\usepackage{cite}
\usepackage{amsmath,amssymb,amsfonts}
\usepackage{algorithm}
\usepackage{algpseudocode}
\usepackage{graphicx} 
\usepackage{bmpsize}
\usepackage{grffile}
\usepackage{subfigure}
\usepackage{subcaption}
\usepackage[belowskip=-10pt,aboveskip=1pt]{caption}
\usepackage{caption}
\usepackage{textcomp}
\usepackage{float}
\usepackage{stfloats}
\usepackage{xcolor}
\usepackage{balance}
\usepackage{dsfont}
\usepackage{multirow}
\usepackage{ntheorem}
\usepackage[colorlinks,
            linkcolor=black,
            anchorcolor=black,
            citecolor=black]{hyperref}
\usepackage{amsmath}
\pdfoutput=1

\ifodd 1
 
\newcommand{\com}[1]{{\color{blue}#1}} 

\else

\newcommand{\com}[1]{}
\fi

\def\BibTeX{{\rm B\kern-.05em{\sc i\kern-.025em b}\kern-.08em
    T\kern-.1667em\lower.7ex\hbox{E}\kern-.125emX}}

\begin{document} 
\title{{Neural Representation for Wireless Radiation Field Reconstruction: A 3D Gaussian Splatting Approach}\\
}
\author{Chaozheng Wen, Jingwen Tong,~\IEEEmembership{Member,~IEEE}, Yingdong Hu, Zehong Lin,~\IEEEmembership{Member,~IEEE}, \\and Jun Zhang,~\IEEEmembership{Fellow,~IEEE}
\thanks{

This work was partly presented at the IEEE International Conference on Computer Communications (INFOCOM) 2025 \cite{wen2025wrf}.
The authors are with the Department of Electronic and Computer Engineering, The Hong Kong University of Science and Technology, Hong Kong (e-mail: cwenae@connect.ust.hk; eejwentong@ust.hk; yhudj@connect.ust.hk; eezhlin@ust.hk; eejzhang@ust.hk). \emph{(Corresponding authors: Jingwen Tong; Zehong Lin.)}}}

\maketitle
\begin{abstract}
Wireless channel modeling plays a pivotal role in designing, analyzing, and optimizing wireless communication systems. Nevertheless, developing an effective channel modeling approach has been a long-standing challenge. This issue has been escalated due to denser network deployment, larger antenna arrays, and broader bandwidth in next-generation networks. To address this challenge, we put forth WRF-GS, a novel framework for channel modeling based on wireless radiation field (WRF) reconstruction using 3D Gaussian splatting (3D-GS). WRF-GS employs 3D Gaussian primitives and neural networks to capture the interactions between the environment and radio signals, enabling efficient WRF reconstruction and visualization of the propagation characteristics. The reconstructed WRF can then be used to synthesize the spatial spectrum for comprehensive wireless channel characterization. 
While WRF-GS demonstrates remarkable effectiveness, it faces limitations in capturing high-frequency signal variations caused by complex multipath effects. To overcome these limitations, we propose WRF-GS+, an enhanced framework that integrates electromagnetic wave physics into the neural network design. WRF-GS+ leverages deformable 3D Gaussians to model both static and dynamic components of the WRF, significantly improving its ability to characterize signal variations. In addition, WRF-GS+ enhances the splatting process by simplifying the 3D-GS modeling process and improving computational efficiency. Experimental results demonstrate that both WRF-GS and WRF-GS+ outperform baselines for spatial spectrum synthesis, including ray tracing and other deep-learning approaches. Notably, WRF-GS+ achieves state-of-the-art performance in the received signal strength indication (RSSI) and channel state information (CSI) prediction tasks, surpassing existing methods by more than 0.7 dB and 3.36 dB, respectively. The code is available at \url{https://github.com/wenchaozheng/WRF-GSplus}.

\end{abstract}

\section{Introduction}\label{SecIntro}
\begin{figure}[t]
    \centering \includegraphics[width=0.48\textwidth]{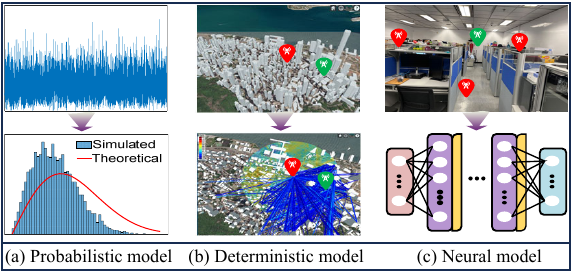}
    \caption{Different types of wireless channel modeling.}
    \label{ChanMod}
\end{figure}

Modern communications increasingly depend on wireless technologies that use electromagnetic waves (EM) for information exchange, driving advances in mobile phones, automotive systems, and the Internet of Things \cite{de2021survey}. At the core of these advances lies \textit{wireless channel modeling}, a long-standing problem for modeling
the interactions between the environment and radio signals in wireless communications. While Maxwell's equations model such interactions \cite{yun2015ray}, solving these equations in reality is intricate due to the need for comprehensive knowledge of boundary conditions. This complexity has led to the development of various wireless channel modeling approaches, as shown in Fig. \ref{ChanMod}, which are categorized into probabilistic modeling, deterministic modeling, and neural modeling. 

The probabilistic models rely on statistical methods to predict channel characteristics, primarily estimating the received signal strength based on the distance between the transmitter (TX) and receiver (RX). These models are based on empirical formulas and use measurements to calibrate parameters for typical scenarios \cite{sarkar2003survey}. However, they often lack accuracy and struggle to provide detailed channel characteristics. For example, in multi-antenna systems, the probabilistic models fail to characterize the energy distribution of the received signal from all directions, i.e., to reconstruct the \textit{spatial spectrum} of the received signal by estimating its angle of arrival (AoA). To overcome these drawbacks, the deterministic models use physical principles to predict channel characteristics under an approximate environmental model. For example, the ray tracing method generates propagation characteristics based on computer-aided design representations of the environment, including object boundaries and material reflection coefficients \cite{he2018design}. Therefore, these models can provide more comprehensive channel information than probabilistic models. However, their accuracy may be compromised, as they cannot capture the detailed physical characteristics of the environment.

In contrast, the neural models, adopting the data-driven principle, learn the complex interactions between the environment and radio signals directly from location-based data. One of the promising advancements in neural models is the recent adoption of the neural radiance field (NeRF), a breakthrough in computer vision for view synthesis \cite{mildenhall2021nerf}. Motivated by the fact that light is a kind of EM wave, Refs. \cite{zhao2023nerf2} and \cite{lunewrf} proposed two NeRF-based frameworks, named NeRF\textsuperscript{2} and NeWRF, respectively, for wireless channel modeling based on implicit \textit{wireless radiation field} (WRF) reconstruction. The reconstructed WRF can provide detailed and accurate channel characteristics and significantly enhance communication performance. However, these methods suffer from high computational complexity and slow synthesis (a.k.a. rendering) speeds. In practice, the synthesis operation corresponds to the channel characteristic prediction process and will significantly affect the round-trip time in communication systems, which is critical for latency-sensitive applications. For instance, cloud gaming and digital twins typically demand low latency, i.e., under 20 milliseconds for cloud gaming \cite{yates2017timely} and often within a few milliseconds for digital twins \cite{10580971}. This renders the NeRF-based methods impractical.

\begin{figure*}[!t]
\centering
\includegraphics[width=1\textwidth]{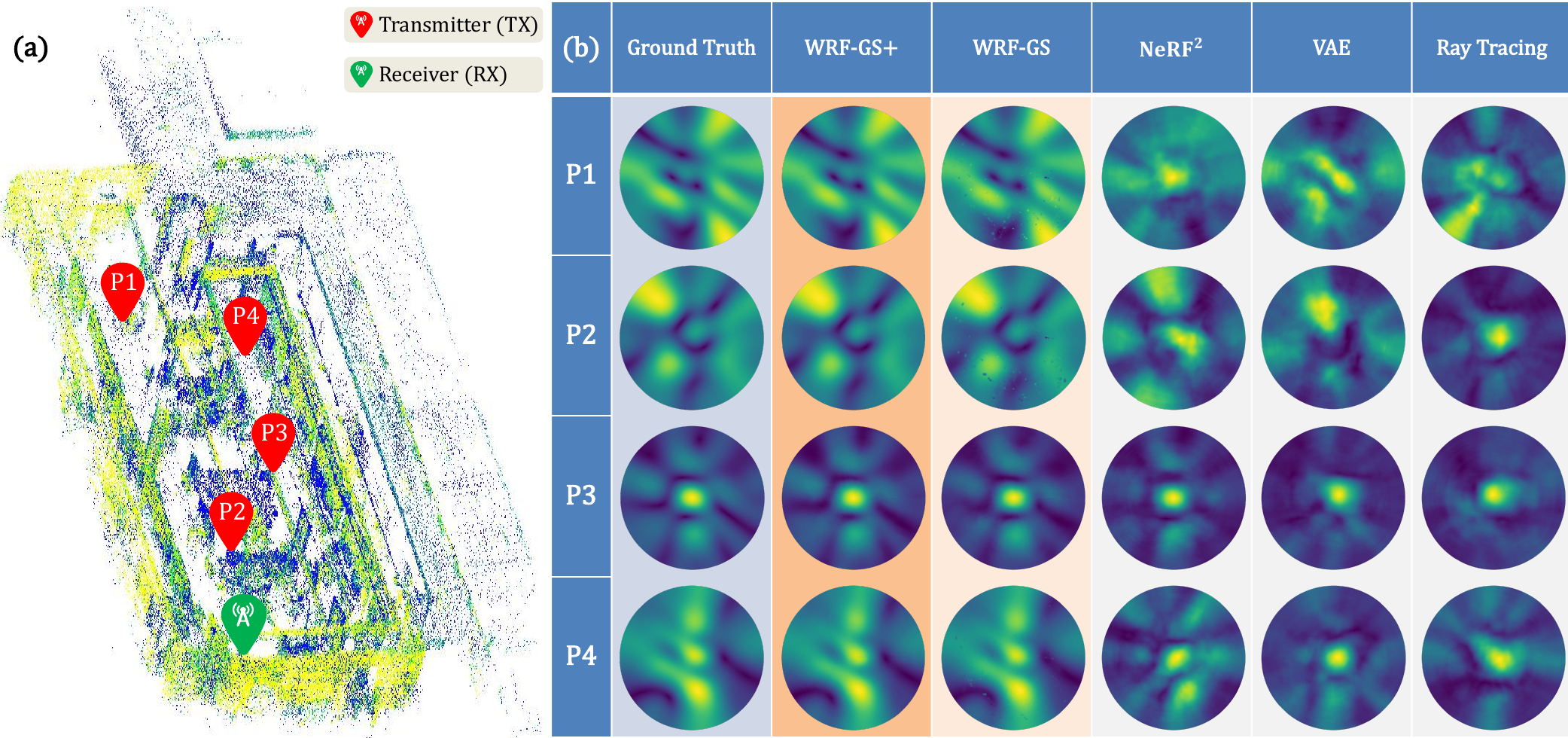}
\caption{The synthesized spatial spectra of five algorithms in a laboratory environment. The spatial spectrum represents the signal strength received from different directions composed of the azimuthal and elevation angles. (a) shows the 3D point clouds of the laboratory created by LiDAR, which are used for the ray tracing algorithm. In this setting, the TX can be located at any position, while the RX is equipped with a $4 \times 4$ antenna array and fixed at a corner.
(b) compares the synthesized spatial spectra of five algorithms when the TX is located at four different positions, i.e., P1-P4, as shown in (a). The ground truth is obtained using the antenna array.}\label{SynSS}
\end{figure*}

Recently, 3D Gaussian splatting (3D-GS) has emerged as a promising technique for efficient and compact scene reconstruction using a set of 3D Gaussian functions \cite{kerbl3Dgaussians}. Compared with NeRF, 3D-GS offers significantly faster rendering speeds and lower computational complexity. These advantages motivate us to adapt the 3D-GS technique for reconstructing the WRF. Nevertheless, directly applying the optical 3D-GS to the RF domain poses several significant challenges. First, the optical 3D-GS only considers the amplitude of optical signals (i.e., light intensity). In contrast, the received RF signals consist of both amplitude and phase components, as EM waves are more prone to reflection, diffraction, and scattering than visible light. Second, visible light is typically measured using million-pixel cameras, while an RX employs either a single antenna or an antenna array. Third, the rendering function in the RF domain differs from that in optical 3D-GS due to the unique physics of EM wave propagation. 

To address these challenges, we propose WRF-GS, a novel framework for fast and accurate channel modeling based on explicit WRF reconstruction using neural networks and 3D-GS. Specifically, we refine the optical 3D-GS technique from three key aspects to adapt it to the RF domain. First, we design a scenario representation network that comprises two multilayer perceptions (MLPs) to effectively capture the complex interactions between the environment and radio signals. The learned features are then embedded into 3D Gaussian points, which serve as virtual TXs to model wireless signal propagation. This approach enables the network to represent the spatial distribution of signals. Second, we devise a projection model to adapt the optical camera model to an RF antenna model using spherical and Mercator projections. This model projects the virtual TXs onto the RX perception plane, enabling accurate spatial mapping of signal propagation. Third, we introduce an electromagnetic splatting process using a differentiable tile rasterizer to synthesize wireless channel characteristics based on the projected 3D Gaussian points. This process, combined with the projection model, enables rapid synthesis of new channel characteristics within milliseconds.

However, WRF-GS faces limitations in accurately modeling high-frequency signal variations and also suffers from parameter redundancy due to its coarse neural network design. To overcome these limitations, we further put forth WRF-GS+, an enhanced version of WRF-GS that integrates electromagnetic wave physics into the neural network design and eliminates the need for additional parameters. Specifically, we adopt the deformable 3D Gaussians in the scenario representation network to model both static and dynamic components of the WRF, significantly improving the ability to capture the complex interactions between the environment and radio signals. In addition, we exploit the inherent properties of 3D Gaussian primitives directly and leverage the $\alpha$-blending technique to synthesize wireless channel characteristics based on the projected 3D Gaussian points. These enhancements enable WRF-GS+ to characterize the distribution of the WRF in a given environment more effectively and with greater computational efficiency. Fig. \ref{SynSS} illustrates the synthesized spatial spectra of five algorithms based on the reconstructed WRF in a laboratory environment. Evidently, WRF-GS+ achieves the best prediction accuracy compared with the WRF-GS method and other baselines, closely matching the ground truth.

We conduct extensive experiments to evaluate the WRF-GS and WRF-GS+ methods. The numerical results show that the proposed methods are capable of explicitly representing the propagation characteristics and achieving higher reconstruction accuracy, greater sample efficiency, and faster rendering speed in wireless channel modeling. Moreover, we present two case studies, i.e., the received signal strength indicator (RSSI) prediction and the downlink channel state information (CSI) prediction, to demonstrate the effectiveness of the proposed methods in practical systems. The RSSI prediction task focuses on estimating the signal strength at a specific location, while the CSI prediction task involves estimating the downlink CSI based on the uplink CSI measurements in a multiple-input-multiple-output (MIMO) system. The numerical results show that WRF-GS+ achieves superior performance in both tasks, outperforming existing methods by more than $0.7$ dB in RSSI prediction and $3.36$ dB in CSI prediction.

Our main contributions are summarized as follows:
\begin{itemize}
    \item We propose WRF-GS, a hybrid framework tailored for wireless channel modeling based on WRF reconstruction using neural networks and 3D-GS. This framework enables efficient and accurate modeling of wireless propagation characteristics.
    \item We adapt the optical 3D-GS to the RF domain by designing the modules of scenario representation network, projection model, and electromagnetic splatting. The integration of the EM wave physics and 3D-GS enables a visible, low-cost, and accurate channel modeling. 
    \item We propose WRF-GS+, an enhanced version of WRF-GS, which employs deformable 3D Gaussians to separately model the static and dynamic components of the signals from virtual TXs. This approach improves efficiency and accuracy in characterizing the distribution of WRF.
    \item We conduct extensive experiments and two case studies to demonstrate the effectiveness of the WRF-GS and WRF-GS+ methods. Numerical results show that the proposed methods consistently outperform the existing methods.
\end{itemize}

The remainder of this paper is organized as follows. 
Section \ref{SecRW} reviews the related works.
Section \ref{SecP} provides the preliminaries of the wireless channel model and 3D-GS.
Sections \ref{SecSM} and \ref{SecSMP} detail the proposed WRF-GS and WRF-GS+ methods, respectively.
Section \ref{SecIAE} presents the implementation and evaluation of the proposed methods in a laboratory environment using a real-world dataset.
Two case studies are presented in Section \ref{SecCS} to further demonstrate the practical applicability of the methods. Finally, Section \ref{SecCON} concludes the paper.

\section{Related Works}\label{SecRW}
\textbf{Wireless channel modeling} involves measuring and characterizing the propagation of EM waves between TXs and RXs, taking into account various phenomena such as reflection, diffraction, scattering, and path loss \cite{oestges2002deterministic, tong2018cooperative}. Sarkar \textit{et al.} \cite{sarkar2003survey} surveyed various propagation models for mobile communications using statistical methods to describe the random nature of EM wave propagation. To provide more detailed channel characteristics, Oestges \textit{et al.} \cite{oestges2002deterministic} applied the ray tracing technique to simulate the paths of EM waves, which allows for precise predictions and accounts for large-scale features of the environment. These methods, however, fail to capture the materials and physical characteristics of the environment.

\textbf{Environment-aware channel modeling} has attracted much attention since future wireless networks will generate abundant location-specific channel data and are expected to possess more powerful data mining and AI capabilities \cite{zeng2024tutorial, letaief2019roadmap}. This motivated the adaptation of neural networks for more accurate channel modeling, i.e., learning the complex interactions between the environment and radio signals directly from location-based data \cite{zhang2024wisegrt, chen2024rfcanvas}. To achieve this goal, a few recent works have been devoted to NeRF-based wireless channel modeling \cite{orekondy2023winert, zhao2023nerf2, lunewrf}. Orekondy \textit{et al.} \cite{orekondy2023winert} proposed a neural surrogate to model wireless EM propagation effects in indoor environments based on NeRF using two synthetic datasets. Meanwhile, Zhao \textit{et al.} \cite{zhao2023nerf2} presented NeRF\textsuperscript{2} for spatial spectrum reconstruction in a real-world environment. By fixing a TX at an indoor location, Lu \textit{et al.} \cite{lunewrf} presented NeWRF, a NeRF-based wireless channel prediction framework with a sparse set of channel measurements. However, NeRF-based methods often suffer from high computational complexity and slower synthesis speed. In this paper, we present WRF-GS for environment-aware channel modeling with 3D-GS, reconciling the accuracy and computational complexity.

\textbf{RSSI prediction} is a typical task in wireless networks to estimate the signal strength at a specific location based on factors such as distance, environment, and prior measurements. This task is crucial for network management, indoor positioning, and signal coverage planning.  Shin \textit{et al.} \cite{shin2014mri} proposed an autonomous smartphone-based war-walking system (MRI) that automatically constructs a radio fingerprint database using limited user-collected data. MRI uses a log-distance path loss model to predict RSSI values throughout the entire indoor space based on partial measurements. Similarly, Zhao \textit{et al.} \cite{zhao2023nerf2} devised a turbo learning-enabled RSSI prediction method based on the NeRF\textsuperscript{2} framework using the Bluetooth Low Energy (BLE) dataset. In this paper, we apply the WRF-GS and WRF-GS+ methods to the RSSI prediction task by retraining them on the public BLE dataset \cite{zhao2023nerf2}.

\textbf{CSI prediction} involves estimating future channel conditions based on current and historical data. Existing studies have focused on predicting downlink channel states by observing uplink channels, as both are influenced by the same underlying physical environment and traverse identical paths \cite{2021Liu, vasisht2016eliminating, bakshi2019fast}. Liu \textit{et al.} \cite{2021Liu} introduced FIRE, a variational auto-encoder (VAE) method, to transfer estimated CSI from the uplink to the downlink. In addition, Vasisht \textit{et al.} \cite{vasisht2016eliminating} and Bakshi \textit{et al.} \cite{bakshi2019fast} proposed R2F2 and OptML to predict the CSI using machine learning-based methods. In this paper, we apply the WRF-GS and WRF-GS+ methods to the CSI prediction task by transferring the estimated CSI from uplink to downlink.

\section{Preliminaries}\label{SecP}
In this section, we present preliminaries on wireless channel modeling and 3D-GS technique.

\subsection{Wireless Channel Modeling} \label{SecP_A}
A generic wireless communication system typically consists of a TX that generates and modulates an information signal, which is then propagated through the wireless channel to an RX. The transmitted signal can be mathematically represented as a complex number $s = Ae^{j\varphi}$, where $A$ and $\varphi$ denote the amplitude and phase of the signal, respectively. In the wireless medium, the propagation of EM signals is influenced by fundamental physical phenomena, including path loss and multipath propagation. In the case of free space path loss, the received signal $y$ at the RX is expressed as:
\begin{equation}
    y = Ae^{j\varphi} \cdot \Delta Ae^{j\Delta \varphi},
    \label{eq:complex_value}
\end{equation}
where $\Delta A$ and $\Delta \varphi$ are the amplitude attenuation factor and phase rotation incurred during propagation, respectively. However, in realistic environments, the signal encounters a multitude of propagation effects, including reflection, scattering, refraction, and diffraction. These phenomena cause the transmitted signal to split into multiple replicas, each traversing different paths before arriving at the RX. As a result, the received signal is more accurately modeled as the superposition of these multipath components:
\begin{equation}
    y = Ae^{j\varphi}  \sum_{l=0}^{L-1} \Delta A_l e^{j\Delta \varphi_l},
    \label{eq:attenuation}
\end{equation}
where $L$ denotes the total number of propagation paths, and $\Delta A_l$ and $\Delta \varphi_l$ correspond to the amplitude attenuation factor and phase rotation specific to the $l$-th path, respectively. This multipath formulation captures the complex interplay of signal interactions with the environment, providing a comprehensive framework to characterize the behaviors of wireless channels.

\begin{figure}[!t]
    \centering
    \includegraphics[width=0.48\textwidth]{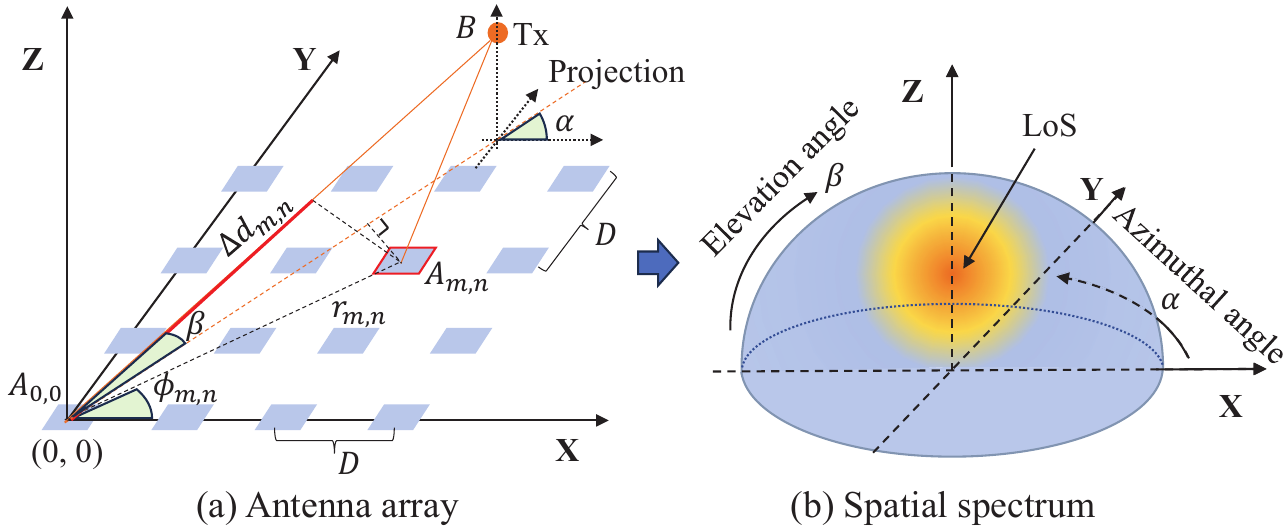}
    \caption{The AoA computation of the received signal with an antenna array. (a) There are $K$ antennas deployed in a grid, and the signal source is located at point B. (b) The spatial spectrum is generated from the antenna array. The illustration of the spatial spectrum in Fig. \ref{SynSS}(c) is obtained by projecting the spatial spectrum to the X-Y plane.}
    \label{AntenMod}
\end{figure}

In multi-antenna systems, an antenna array is employed to characterize the spatial energy distribution of the received signal $y$ from all directions, i.e., to estimate the AoA. To illustrate this concept, we consider an antenna array equipped with $\sqrt{K} \times \sqrt{K}$ antennas, as depicted in Fig. \ref{AntenMod}. The spacing of two adjacent antennas is $D$, where $D < \lambda$ and $\lambda$ denotes the wavelength of the transmitted signal. As shown in Fig. \ref{AntenMod}(b), the antenna array operates within a hemispherical plane, providing omnidirectional coverage to efficiently capture signals from various directions. The direction of an RF source is characterized by two angles: the azimuthal angle $\alpha$ $(0^{\circ} \leq \alpha < 360^{\circ})$ and the elevation angle $\beta$ $(0^{\circ} \leq \beta < 90^{\circ})$. The phase difference between the signals received by the $(m,n)$-th antenna pair is determined by the relative distance of the two antennas from the source \cite{an2020general}, expressed as:
\begin{equation*}
    \begin{aligned}
\Delta \theta_{m, n} &= \mathrm{mod} \left(\frac{2\pi \Delta d_{m,n}}{\lambda}, 2 \pi \right)\\
&= \mathrm{mod} \left(\frac{-2 \pi r_{m, n} \cos \left(\alpha-\phi_{m, n}\right) \cos (\beta)}{\lambda}, 2 \pi \right),
\end{aligned}
\end{equation*}
where $\Delta d_{m,n}$, illustrated in Fig. \ref{AntenMod}(a), is computed as
\begin{equation*}
    \Delta d_{m,n} = |BA_{m,n}|-|BA_{0,0}| = -r_{m,n}\cos(\alpha-\phi_{m,n})\cos \beta, 
\end{equation*}
Here, $m, n = 0, \ldots ,\sqrt{K} - 1$, $r_{m,n} = D\sqrt{m^2+n^2}$ represents the radial distance of antenna $A_{m,n}$ in a polar coordinate system, and $\phi_{m,n} = \arctan 2(n,m)$ is its angular position.

The AoA of the received signal is calculated by comparing the phase differences of the signals received across multiple antennas. To achieve this, we use the antenna array to form a narrow beam and steer it across different directions. When the beam aligns with the line-of-sight (LoS) direction, the array effectively suppresses signals from other directions. This allows us to reconstruct the spatial spectrum by steering the beam across a grid of azimuth and elevation angles and measuring the received signal power at each pointing direction. Specifically, we can form a matrix $\mathbf{P}$, where each element $\mathrm{P}(\alpha, \beta)$ represents the received signal power from the azimuthal angle $\alpha$ and elevation angle $\beta$, i.e.,
\begin{equation}
\mathrm{P}(\alpha, \beta)
=\left|\frac{1}{K} \sum_{m, n=0}^{\sqrt{K}-1, \sqrt{K}-1} e^{{j}\left(\Delta \hat{\theta}_{m, n}-\Delta \theta_{m, n}\right)}\right|^{2}, \\
\label{eq:ss01}
\end{equation}
with
\begin{equation*}
 \Delta \hat{\theta}_{m, n} = \hat{\theta}_{m, n} - \hat{\theta}_{0, 0},   
\end{equation*}
where $\hat{\theta}_{m, n}$ denotes the measured phase of the RF signal received at antenna $A_{m,n}$, and $\hat{\theta}_{0,0}$ is a constant reference phase. Considering a one-degree angular resolution, the spatial spectrum is reconstructed as a $360\times 90$ matrix:
{\small \begin{align}
\resizebox{0.43\textwidth}{!}{$
    \mathbf{P}=\begin{bmatrix}
  \mathrm{P}(0^\circ, 0^\circ )  &\mathrm{P}(1^\circ, 0^\circ )  &\cdots  &\mathrm{P}(359^\circ, 0^\circ ) \\
  \mathrm{P}(0^\circ, 1^\circ )  &\mathrm{P}(1^\circ, 1^\circ ) &\cdots  &\mathrm{P}(359^\circ, 1^\circ )  \\
  \vdots &\vdots &\ddots &\vdots  \\
    \mathrm{P}(0^\circ, 89^\circ ) &\mathrm{P}(1^\circ, 89^\circ ) &\cdots  &\mathrm{P}(359^\circ, 89^\circ) 
\end{bmatrix}$}.
\label{eq:ss02}
\end{align}}

The reconstructed spatial spectrum represents a spatial power distribution function, which directly corresponds to the WRF emitted by the transmitting source and propagated through the environment. In other words, the spatial spectrum serves as a representation of the underlying WRF, providing valuable insights into wireless channel modeling. Therefore, this paper aims to reconstruct the WRF and synthesize the spatial spectrum using the 3D-GS technique, which is introduced in the following subsection.

\subsection{3D Gaussian Splatting} \label{SecP_B}
3D-GS is a technique that utilizes a set of 3D Gaussian functions to model the geometry and appearance of a scene, offering a compact and efficient representation \cite{kerbl3Dgaussians, chen2024gi}. 
Mathematically, 3D-GS represents space using anisotropic Gaussian primitives, where each Gaussian is characterized by its covariance matrix $\mathbf{\Sigma} \in \mathbb{R}^{3 \times 3 }$ and center position vector $\boldsymbol{\mu} \in \mathbb{R}^{3}$. The 3D Gaussian distribution is expressed as
\begin{equation}
    G ( \boldsymbol{x} )= {e}^{-\frac{1}{2}(\boldsymbol{x}-\boldsymbol{\mu} )^{T} \boldsymbol{\Sigma}^{-1} (\boldsymbol{x}-\boldsymbol{\mu}) },
\end{equation}
where $\boldsymbol{x} = (x_{0},x_{1},x_{2})$ represents the spatial coordinates of a Gaussian point in 3D space, and $\boldsymbol{\Sigma}$ can be expressed in terms of a scaling matrix $\mathbf{S}$ and a rotation matrix $\mathbf{R}$ as
\begin{equation}
    \mathbf{\Sigma} =\mathbf{R}\mathbf{S}\mathbf{S}^{T}\mathbf{R}^{T}.
\end{equation}

  The rendering process begins by projecting the 3D Gaussians $G(\boldsymbol{x})$ onto the image plane, resulting in 2D Gaussians $G'(\boldsymbol{x'})$  \cite{zwicker2002ewa}. These 2D Gaussians are then sorted based on their depth information to ensure proper occlusion handling. Each tile on the image plane is processed independently and in parallel for the 2D Gaussians within its coverage. Using the opacity ${o}_{i}$ and color attribute $c_i$ of each Gaussian, $\alpha$-blending is applied to compute the color values for each pixel:
\begin{equation}
    C = \displaystyle\sum_{i=1}^{N}{c}_{i}{\alpha}_{i}\displaystyle\prod_{j=1}^{i-1}(1-{\alpha}_{j}),
\label{eq:redering}
\end{equation}
where
\begin{equation*}
    {\alpha}_{i}={o}_{i}{G}_{i}^{'}(\varDelta \boldsymbol{p}_{i}).
\end{equation*}
Here, $N$ denotes the number of 3D Gaussians affecting a given pixel, and $\varDelta \boldsymbol{p}_{i}=\boldsymbol{p}_{i}-\boldsymbol{p}_{s}$ represents the difference vector between the projected center $\boldsymbol{p}_{i}$ of a Gaussian and the sampling pixel position $\boldsymbol{p}_{s}$. 
Once the color of each pixel is computed, the entire image is rendered. The rendered image is then compared with the ground truth to calculate the pixel-wise loss, which is used to optimize the model parameters.

In summary, 3D-GS is an efficient and scalable technique for representing and rendering 3D scenes in computer graphics by modeling scenes as a collection of Gaussian primitives. This motivates us to adopt the 3D-GS technique for wireless channel modeling based on explicit WRF reconstruction. This adaptation is achieved by treating Gaussian primitives as virtual TXs that capture wireless signal strengths and attenuation properties. By mapping the learned Gaussian representations onto the receiver's hemispherical antenna perception plane, we can efficiently reconstruct the WRF and rapidly synthesize accurate spatial power spectra in Eqn. \eqref{eq:ss02} and Fig. \ref{SynSS}(b).

\section{The WRF-GS Framework}\label{SecSM}
In this section, we present the WRF-GS framework, a novel approach for wireless channel modeling based on the reconstruction of WRF using 3D-GS. This framework leverages the power of 3D-GS to explicitly characterize the spatial distribution of the WRF in a given environment. The key innovation lies in transforming spatial particles into virtual TXs using a continuous volumetric scene representation enabled by 3D-GS.

\begin{figure*}[!t]
    \centering
    \includegraphics[width=1\textwidth]{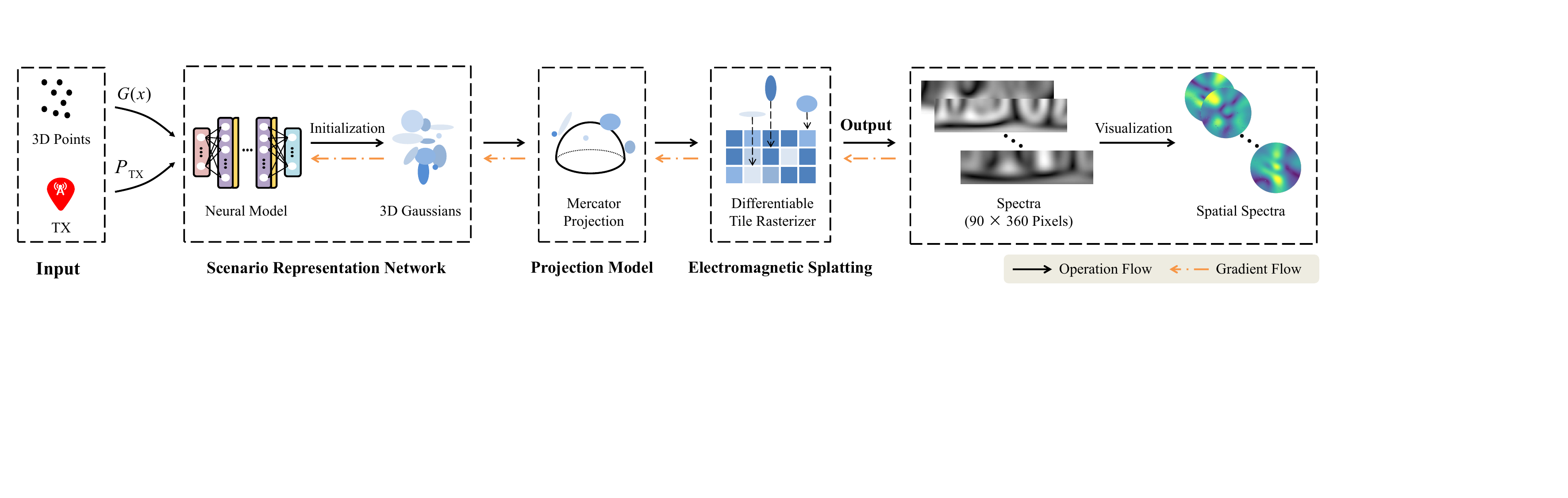}
    \caption{An overview of the WRF-GS framework. The 3D points, which can be randomly distributed or captured via LiDAR sensors, and the position of the TX are first passed into a scenario representation network. This network represents the virtual TXs in the scene using a set of 3D Gaussians, each of which carries environmental attenuation information and signal characteristics. To project these 3D Gaussian representations onto the perception plane of the RX antenna array, the Mercator projection is employed. Finally, the resulting spatial spectra are rendered using the electromagnetic splatting method.}
    \label{Framework}
\end{figure*}
\subsection{Problem Description and Framework Overview} \label{problem_description}
We consider a wireless communication scenario comprising a TX, an RX equipped with an antenna array, and various obstacles such as floors, walls, and furniture. The RX is positioned at a fixed location, while the TX is randomly placed within the environment. Following \cite{zhao2023nerf2}, we assume that the primary obstacles in the environment remain stationary, and any transient perturbations caused by moving small obstacles are mitigated using techniques such as Kalman filtering to minimize their impact. As described in Section \ref{SecP_A}, the received signal at the RX is a superposition of sub-signals from multiple propagation paths. Each sub-signal can be represented as a LoS signal originating from a virtual TX. Our goal is to reconstruct the spatial power spectrum received at the RX antenna array by modeling the distribution of these virtual TXs within the environment.

This task is analogous to the optical 3D-GS technique, which explicitly models the distribution of particles in a scene to enable arbitrary view synthesis \cite{kerbl3Dgaussians}. This motivates us to adapt 3D-GS for WRF reconstruction and propose the novel WRF-GS framework. Specifically, WRF-GS models the distribution of virtual TXs to synthesize the received spatial spectrum by integrating the EM wave physics with 3D-GS. Since 3D Gaussians possess attributes such as volume, color, and opacity, serving as the fundamental building blocks in the representation space, they can be naturally reinterpreted as virtual TXs in the RF domain. During EM propagation, the overall signal is perceived as the combined effect of signal strength and attenuation, as described in Eqn. \eqref{eq:attenuation}. In this context, the color attribute of a Gaussian corresponds to the signal strength, while the opacity represents the attenuation. 

However, the projection and splatting processes for WRF reconstruction in 3D-GS differ significantly from those in the optical domain, as described in Section \ref{SecP_B}. In optical 3D-GS, the goal is to project 3D Gaussian particles representing objects onto a 2D image plane, which involves a coordinate projection transformation based on the camera's position, orientation, and intrinsic parameters. In contrast, the RF domain requires mapping the virtual TXs represented by 3D Gaussians onto the perception plane of the RX antenna array. Since the RX antenna array receives signals in a hemispherical plane, as discussed in Section \ref{SecP_A}, a new projection approach is needed to map 3D Gaussians onto a hemispherical plane rather than a flat 2D plane. This operation is critical for aligning the virtual TXs with the specific directional sensitivity of the RX antenna array. In addition, while optical 3D-GS renders images by blending the volume, color, and opacity of splatted Gaussians, the RF domain requires blending the signal strengths and attenuations of virtual TXs to synthesize the received spatial power spectrum. This necessitates a new splatting process tailored to the properties of EM wave propagation.

Our proposed WRF-GS framework effectively addresses the aforementioned challenges and enables the accurate reconstruction of the spatial spectrum. The overall architecture of WRF-GS is illustrated in Fig. \ref{Framework}. It takes as input the positions of physical TXs and a set of randomly initialized 3D points representing spatial particles within the environment, and outputs the spatial spectrum at the RX antenna array. WRF-GS consists of the following three core modules:
\begin{itemize}
    \item \textbf{Scenario Representation Network:} This module represents the virtual TXs using 3D Gaussians. By processing the input data, the 3D Gaussians capture the signal strength and attenuation properties, enabling the synthesis of the WRF in the environment.

    \item \textbf{Projection Model:} This model projects the virtual TXs represented by 3D Gaussians onto the perception plane of the RX antenna array. To account for the hemispherical nature of the antenna's reception, we employ the Mercator projection to achieve this spatial mapping.

    \item \textbf{Electromagnetic Splatting:} Leveraging the properties of 3D Gaussians and the characteristics of EM wave propagation, this module employs a hardware-accelerated algorithm to efficiently combine the contributions of the virtual TXs onto the RX antenna array, ultimately synthesizing the received spatial spectrum.
\end{itemize}

In the following, we elaborate on these three modules.

\subsection{Scenario Representation Network}
\begin{figure}[!t]
    \centering
    \includegraphics[width=0.48\textwidth]{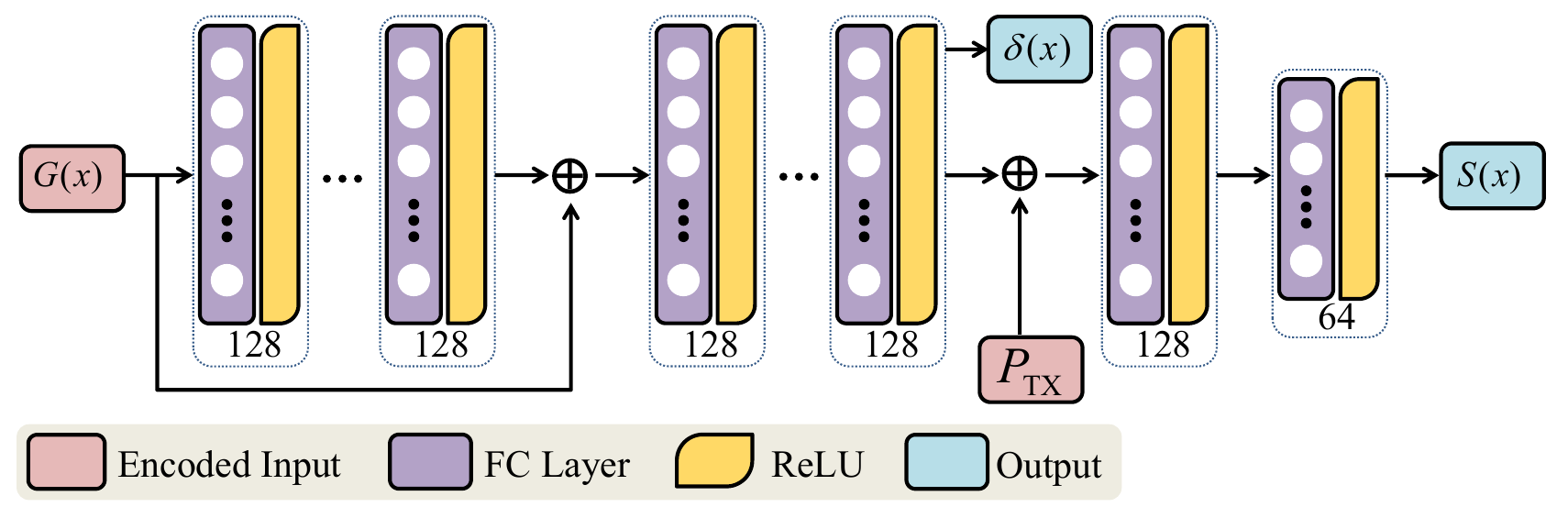}
    \caption{Architecture of the neural model. The model comprises two MLPs. The first MLP processes the input 3D point clouds to capture the spatial attenuation information of the environment, which is independent of the specific TX position. This MLP outputs the attenuation and a feature vector for each spatial location. The second MLP then combines the spatial feature vector with the TX position to capture the signal characteristics.}
    \label{MLP}
\end{figure}

The scenario representation network plays an important role in modeling the intricate relationships between the environment and radio signals by representing virtual TXs as 3D Gaussians using two MLPs. The network architecture, depicted in Fig. \ref{MLP}, is inspired by the DeepSDF structure \cite{park2019deepsdf} but tailored for wireless channel modeling. 

The network begins by processing the initial 3D points, which represent randomly sampled spatial particles within the environment. These points serve as the center coordinates of the 3D Gaussians $G(\boldsymbol{x})$ at location $\boldsymbol{x}$ and are used to initialize the Gaussian primitives. These positions are fed into a neural network to capture the attenuation properties, which depend solely on the spatial location. To characterize these properties, we introduce a new attenuation attribute $\delta (\boldsymbol{x})$ for each 3D Gaussian, which quantifies the signal attenuation caused by environmental interactions at position $\boldsymbol{x}$. This neural network consists of eight fully connected layers, each with ReLU activations and 128 channels. The output of this neural network includes the attenuation $\delta (\boldsymbol{x})$ of each input 3D Gaussain $G (\boldsymbol{x})$ and a feature vector that captures environmental information. Next, this feature vector, along with the TX position $P_{\text{TX}}$, is passed through a second two-layer fully connected network. The first layer has 128 channels and the second layer has 64 channels, both employing ReLU activations. The final output is the signal $S(\boldsymbol{x})$, which represents the 3D Gaussian signal associated with the virtual TX at position $\boldsymbol{x}$. Therefore, for each 3D Gaussian $G (\boldsymbol{x})$ representing a virtual TX, the scenario representation network produces the corresponding signal $S(\boldsymbol{x})$ and attenuation $\delta (\boldsymbol{x})$, as expressed by
\begin{equation}
    {F}_{\Theta}:(G(\boldsymbol{x}), P_{\text{TX}})\Rightarrow\left(\delta(\boldsymbol{x}), S(\boldsymbol{x})\right),
\label{F_theta}
\end{equation}
where $\Theta$ denotes the learnable weights of the network. Since signals and attenuation in space are typically expressed as complex numbers, as shown in Eqn. \eqref{eq:complex_value}, the two outputs are expressed as $\delta(\boldsymbol{x})= \Delta A(\boldsymbol{x})e^{j\Delta \psi(\boldsymbol{x})}$ and $S(\boldsymbol{x})= A(\boldsymbol{x})e^{j \psi(\boldsymbol{x})}$, respectively, where $A(\boldsymbol{x})$ represents the amplitude and $\psi(\boldsymbol{x})$ represents the phase.

This network is specifically designed to transform the 3D points (representing spatial particles) and TX position information into the signal and attenuation properties of 3D Gaussians. Unlike DeepSDF, which is typically applied in optic domains and assumes a fixed TX (light source) position, our network is uniquely adapted to handle mobile TXs. Moreover, the outputs are complex-valued signals that incorporate phase information, a critical feature for accurately representing wireless propagation. This capability enables our model to successfully capture both the signal characteristics of virtual TXs and the attenuation properties of the surrounding environment, making it well-suited for wireless channel modeling.

\subsection{Projection Model}

The projection model is responsible for mapping the virtual TXs represented by 3D Gaussians onto the perception plane of the RX antenna array. There are several key differences compared with the camera model used in optical 3D-GS. First, unlike pinhole or fisheye cameras, the RX antenna in our WRF-GS receives signals from a hemispherical direction. Second, we do not account for distortions caused by coordinate transformations, as we are only concerned with the signal at specific angular resolutions and ignore signals between discrete angles. Third, as shown in Eqn. \eqref{eq:ss02}, the arrangement of pixels in the final spatial spectrum differs from conventional photographs, which are designed for human visual perception. Instead, the spatial spectrum is structured to represent signal strength as a function of azimuth and elevation angles.

\begin{figure}[!t]
    \centering
    \subfigure[Antenna coordinate.]{
        \includegraphics[width=0.45\linewidth]{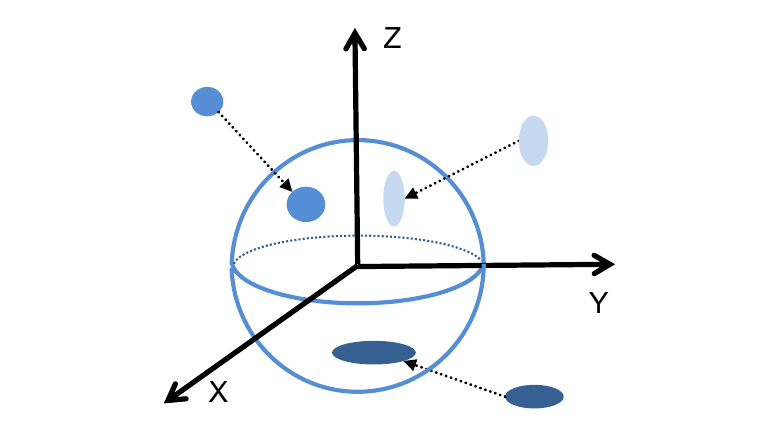}
    }
    \subfigure[Latitude-longitude coordinate.]{
        \includegraphics[width=0.45\linewidth]{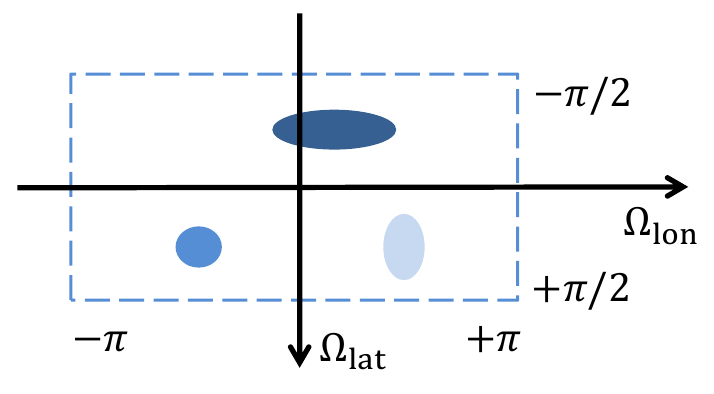}
    }
    \subfigure[Uniform coordinate.]{
        \includegraphics[width=0.45\linewidth]{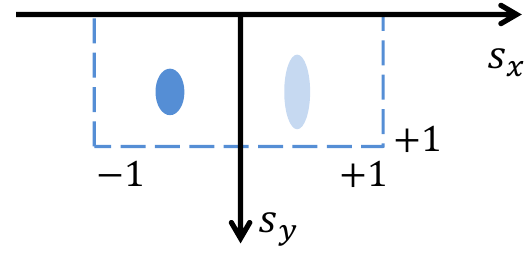}
    }
    \subfigure[Image coordinate.]{
        \includegraphics[width=0.45\linewidth]{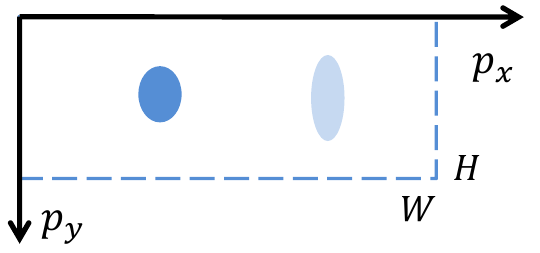}
    }
    \caption{The coordinate projection transformation in the projection model.}
    \label{Proj}
\end{figure}

To project a 3D point $\boldsymbol{t} = {[{t}_{x},{t}_{y},{t}_{z}]}^{T}$ onto the perception plane of the RX antenna array at $\boldsymbol{p}={[{p}_{x},{p}_{y}]}^{T}$, we employ the Mercator projection, a widely used orthographic cylindrical projection in cartography \cite{li2024omnigs}. This allows us to map the latitude and longitude grid of a sphere onto a cylindrical surface, which can then be unrolled onto a flat plane. We define a right-handed Cartesian coordinate system at the receiving end of the antenna, with the antenna array oriented in the positive Z-direction and signals received in a spherical plane, as shown in Fig. \ref{Proj}(a). Using longitude $\Omega_{\text{lon}}$ and latitude $\Omega_{\text{lat}}$, we can concisely and accurately represent positional relationships in space. By applying inverse trigonometric functions, we obtain
\begin{equation}
    \begin{bmatrix}
         
        \Omega_{\text{lon}}\\
        \Omega_{\text{lat}}
        \end{bmatrix}=\begin{bmatrix}
         
        \arctan2({t}_{y}/{t}_{x}) \\
        \arcsin({t}_{z}/{t}_{r})
    \end{bmatrix},
\end{equation}
where ${t}_{r}=\sqrt{{t}_{x}^2+{t}_{y}^2+{t}_{z}^2} $ denotes the distance from the center coordinate of the 3D Gaussian in space to the origin of the antenna coordinate system. In addition, the function $\arctan2(\cdot)$ is the 4-quadrant inverse tangent. Since we only consider signals from the upper hemisphere, the longitude and latitude ranges are constrained to $-\pi \le \Omega_{\text{lon}} < \pi$ and $0 \le \Omega_{\text{lat}} < \pi /2$. Signals from the lower hemisphere are ignored in subsequent coordinate transformations. Next, we convert the latitude-longitude coordinates (see Fig. \ref{Proj}(b)) into uniform coordinates (see Fig. \ref{Proj}(c)) as follows:
\begin{equation}
    \begin{bmatrix}
         {s}_{x}\\
        {s}_{y}
        \end{bmatrix}=\begin{bmatrix}
         \Omega_{\text{lon}}/ \pi\\
         2\Omega_{\text{lat}}/ \pi
    \end{bmatrix},
\end{equation}
such that $-1\le{s}_{x}<1$ and $0 \le  {s}_{y} < 1$. The final step of the projection is to transform the uniform coordinates into pixel coordinates of arbitrary resolution (see Fig. \ref{Proj}(d)), i.e., 
\begin{equation}
    \begin{bmatrix}
         {p}_{x}\\
        {p}_{y}
        \end{bmatrix}=\begin{bmatrix}
         ({s}_{x}+1) \times W/2\\
         {s}_{y} \times H
    \end{bmatrix},
\end{equation}
where $W$ and $H$ represent the number of pixels contained in the width and height of the image, respectively. For a one-degree angular resolution, we set $W=360$ and $H=90$.

Through the above coordinate transformations, we maintain the relationship between the signal angles, ensuring that each pixel value in the spatial spectrum corresponds to the AoA of the signal. This allows us to calculate the signal strength arriving from each direction at the RX antenna array, as detailed in the following subsection.

\subsection{Electromagnetic Splatting}
\label{subsec:ElecSplat}

After the coordinate projection transformation, the 3D Gaussians representing the virtual TXs are mapped onto a 2D plane. Note that each 2D Gaussian may cover multiple pixels. In other words, the value of each pixel is determined by the combined contributions of multiple overlapping 2D Gaussians. The role of the electromagnetic splatting module is to efficiently compute the value of each pixel on the 2D plane, which is then used to reconstruct the spatial spectrum.

As illustrated in Fig. \ref{splatting}, electromagnetic splatting incorporates the principles of EM wave propagation into a differentiable rasterisation algorithm \cite{kerbl3Dgaussians}. To enable fast parallel computing, we group adjacent pixels into non-overlapping ``tiles", which are processed independently, as depicted in Fig. \ref{splatting}(a). For each 2D Gaussian, we first identify the tiles it covers, duplicate the Gaussian, and assign these copies to the corresponding tiles. The 2D Gaussians within each tile are then sorted by depth, as shown in Fig. \ref{splatting}(b).

In the scenario representation network, we obtain the signal $S(\boldsymbol{x}_{i})$ and signal attenuation $\delta(\boldsymbol{x}_{i})$ of each 3D Gaussian $G(\boldsymbol{x}_i)$ at an arbitrary position $\boldsymbol{x}_i$. Assuming there are a total of $N$ 3D Gaussians contributing to a specific angle (pixel), we can express the effect of the $i^{th}$ Gaussian $G(\boldsymbol{x}_i)$ on the RX by combining Eqn. \eqref{eq:attenuation} and Eqn. \eqref{eq:redering} as
\begin{equation}
    S_{i}(\boldsymbol{x}) = \left(\prod_{j=0}^{i-1} \delta(\boldsymbol{x}_{j}) \right) S(\boldsymbol{x}_{i}).
    \label{eq:12}
\end{equation}
Consequently, for all 3D Gaussians associated with a specific angle (pixel) $k$ , the combined effect is expressed as
\begin{equation}
    R_{k}=\sum_{i=1}^{N} S_{i}(\boldsymbol{x}). \label{pixel_val}
\end{equation}

To facilitate a more intuitive understanding of this process, we provide an illustrative example in Fig. \ref{splatting}(c). For the pixel value $R_{1}$ in the upper-left corner, only the signal $S(\boldsymbol{x}_{1})$ from $G(\boldsymbol{x}_{1})$ contributes, without any attenuation. For the pixel value $R_{2}$ in the upper-right corner, the contributions from $G(\boldsymbol{x}_{1})$, $G(\boldsymbol{x}_{2})$, and $G(\boldsymbol{x}_{3})$ must be considered. The signal from $G(\boldsymbol{x}_{1})$ experiences no attenuation, while the signals from $G(\boldsymbol{x}_{2})$ and $G(\boldsymbol{x}_{3})$ are attenuated by $\delta(\boldsymbol{x}_{1})$ and $\delta(\boldsymbol{x}_{1})\delta(\boldsymbol{x}_{2})$, respectively, due to the depth ordering. The computation for the remaining pixel values $R_{k}$ follows the same principle.

In simple terms, we consider each projected Gaussian as a virtual TX that can affect multiple angles, while signals from multiple virtual TXs are received at each angle. Depending on their distance to the RX, these signals undergo varying degrees of attenuation, ultimately determining the received signal strength at each angle. As explained in Section \ref{SecP_A}, the reconstructed spatial spectrum can be represented as a spatial power distribution function, where the value is proportional to the square of the amplitude of the received signal. By deriving the received signal for each angle using Eqn. \eqref{eq:12} and Eqn. \eqref{pixel_val}, we obtain the spatial spectrum described in Eqn. \eqref{eq:ss02}, which serves as a representation of the underlying WRF.

\subsection{Summary}
The WRF-GS framework integrates physical models and deep learning techniques to enable comprehensive environment-aware channel modeling and high-fidelity spatial spectrum synthesis. At the core of this approach lies the establishment of a connection between virtual TXs and 3D Gaussians. The scenario representation network models the spatial distribution of signal propagation and attenuation using 3D Gaussians based on input data. To transform these 3D spatial representations onto a 2D plane, we employ the Mercator map projection, ensuring accurate spatial mapping. Finally, the electromagnetic splatting module employs a parallel and efficient algorithm to compute pixel values for each direction, enabling rapid synthesis of the spatial spectrum.

\section{The WRF-GS+ Framework}\label{SecSMP}
The proposed WRF-GS framework demonstrates remarkable effectiveness in reconstructing the WRF and synthesizing spatial spectra for wireless channel modeling, as shown in Section \ref{SecIAE} and Section \ref{SecCS}. However, WRF-GS faces limitations in accurately modeling high-frequency signal variations, as the static nature of the 3D Gaussians in WRF-GS restricts its ability to capture rapid changes in the channel caused by complex multipath effects. To address this challenge, we propose WRF-GS+, an enhanced framework that introduces deformable 3D Gaussians to better model dynamic signal variations and incorporates electromagnetic splatting with $\alpha$-blending to simplify the modeling process.

\subsection{Scenario Representation Network Using Deformable 3D Gaussians}

While WRF-GS effectively models wireless environments, it struggles to accurately capture the rapid signal variations caused by complex multipath effects. To address this limitation, we introduce deformable 3D Gaussians, which enables the framework to decompose the signal into static components and dynamic components. This decomposition allows the network to separately model large-scale fading (e.g., path loss) and small-scale fading (e.g., multipath effects), significantly improving its ability to handle high-frequency signal changes.

\begin{figure}[!t]
    \centering
    \includegraphics[width=0.40\textwidth]{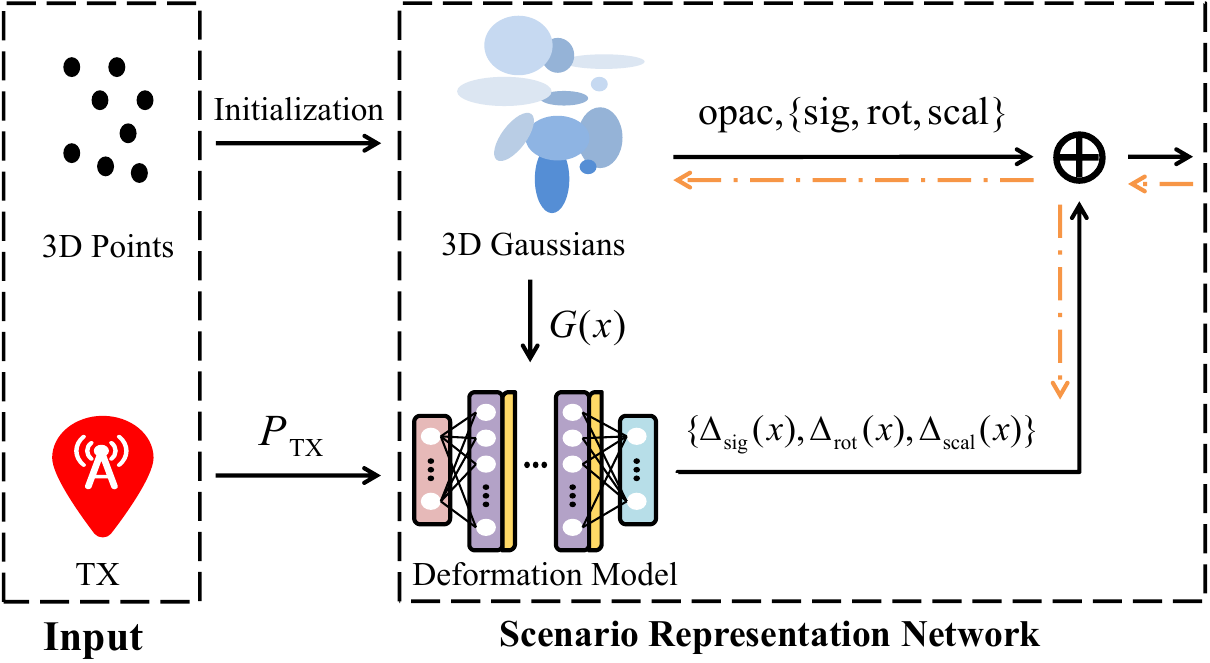}
    \caption{The enhanced scenario representation network. The random 3D points are used to initialize 3D Gaussians, each characterized by properties such as position, opacity, signal strength, rotation, and scale. The 3D Gaussians, along with the TX position, are fed into the deformation network to compute dynamic offsets for the properties. These dynamic offsets are combined with the original static properties to derive the final characteristics of 3D Gaussians. These updated properties are then processed through the projection model and electromagnetic splatting to synthesize the final spatial spectra.} 
    \label{ArcNew}
\end{figure}

As illustrated in Fig. \ref{ArcNew}, the scenario representation network initializes Gaussian primitives from 3D points, each assigned with attributes such as opacity, signal strength, rotation, and scaling. Unlike WRF-GS, which introduces an additional attribute to explicitly model signal attenuation, WRF-GS+ directly leverages the opacity attribute to capture attenuation characteristics. The signal strength, rotation, and scaling attributes are determined exclusively by the center position of the 3D Gaussian. Notably, these static components of the signal, corresponding to large-scale fading, depend solely on the relative position of the virtual TX to the RX, as the environmental distribution remains invariant. These attributes are thus used to characterize the static components: signal strength is represented by the original color attribute, while rotation and scaling define the size and orientation of Gaussians, thereby affecting the coverage area of the virtual TXs.

\begin{figure}
    \centering
    \includegraphics[width=1\linewidth]{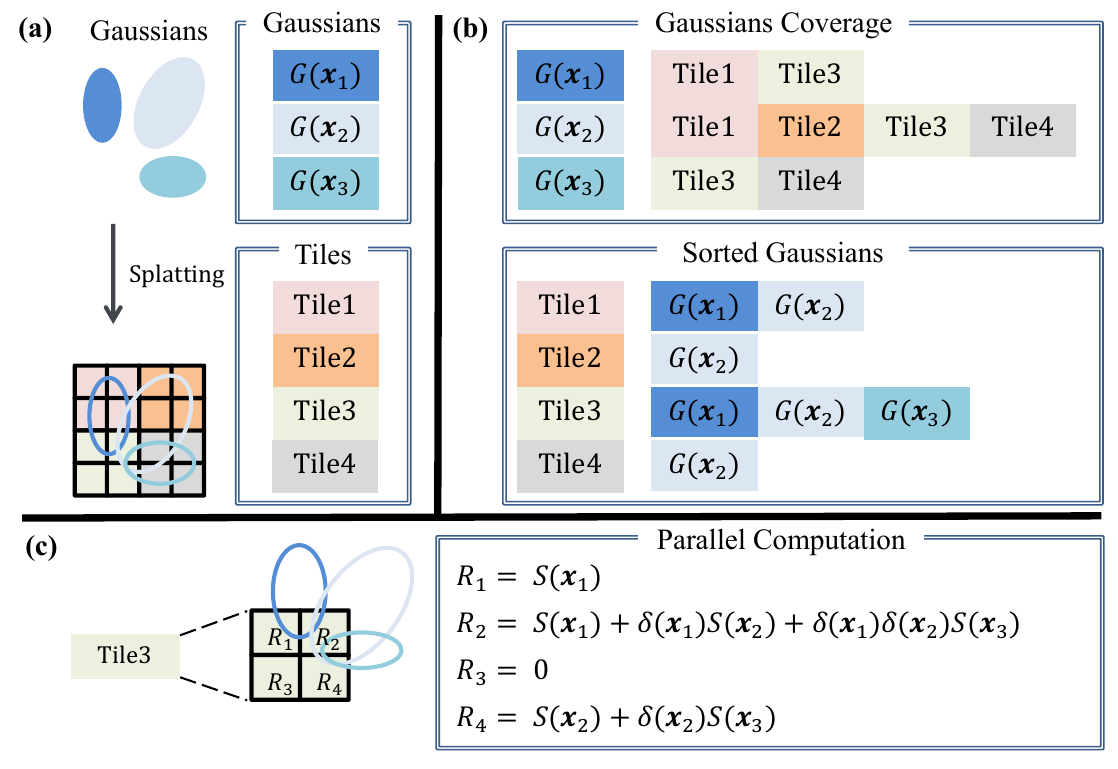}
    \caption{An illustration of electromagnetic splatting. (a) The 3D Gaussians are splatted onto a 2D plane, and the plane is divided into non-overlapping tiles to enable parallel computation. (b) Gaussians that cover multiple tiles are duplicated and assigned to the corresponding tiles, then sorted by depth order within each tile. (c) The value of each pixel can be calculated independently and in parallel using Eqn. \eqref{eq:12} and Eqn. \eqref{pixel_val}.}
    \label{splatting}
\end{figure}

\begin{figure}[!t]
    \centering
    \includegraphics[width=0.48\textwidth]{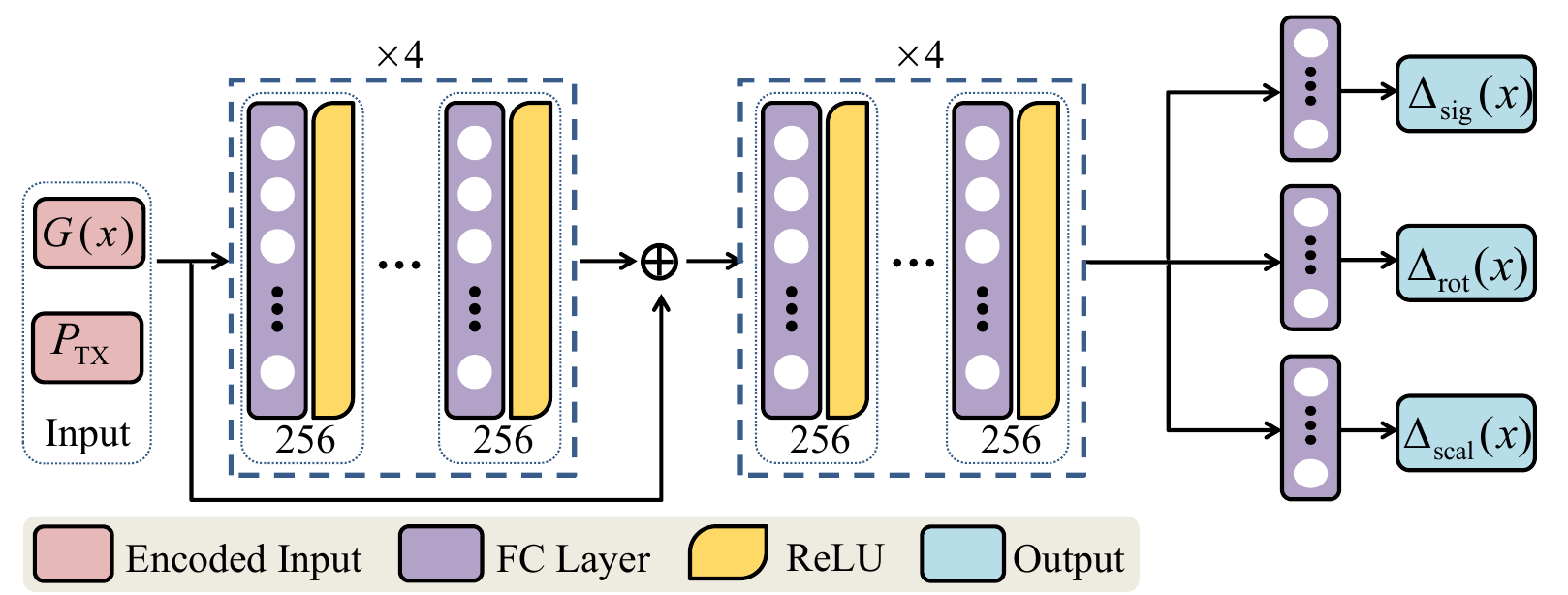}
    \caption{ The architecture of the deformation network, designed to learn the dynamic components in the scene. The 3D Gaussians and the TX position are first encoded using positional encoding and then fed into the deformation network. The network outputs the offsets in signal strength, rotation, and scaling attributes for each 3D Gaussian, enabling the representation of dynamic signal variations caused by TX movement and environmental interactions.}
    \label{DeformMLP}
\end{figure}

The dynamic components, which correspond to small-scale fading, are modeled using a deformation network ${D}_{\Theta}$. This network captures the variations in signal propagation paths caused by complex multipath effects, leading to changes in the coverage and shape of virtual TXs. The deformation network, implemented as an MLP, maps the coordinates of the 3D Gaussians and the TX position to their corresponding deviations in signal strength, rotation, and scaling. The weights $\Theta$ of the MLP are optimized through this mapping process. As shown in Fig. \ref{DeformMLP}, the deformation network processes the input through eight fully connected layers with ReLU activations and 256-dimensional hidden layers, producing a 256-dimensional feature vector. This vector is then passed through three additional fully connected layers (without activation functions) to separately output the offsets for signal strength, rotation, and scaling. Notably, similar to NeRF, we concatenate the feature vector output from the fourth layer with the input and feed them into the fifth layer to enhance feature representation. Given the TX position, it is fed into the network along with the 3D Gaussians to generate the offsets:
\begin{equation}
    {D}_{\Theta}:(G(\boldsymbol{x}), P_{\text{TX}})\Rightarrow\left(  \Delta_{\text{sig}}(\boldsymbol{x}), \Delta_{\text{rot}}(\boldsymbol{x}),  \Delta_{\text{scal}}(\boldsymbol{x})   \right),
\label{D_theta}
\end{equation}
where $\Delta_{\text{sig}}(\boldsymbol{x})$, $\Delta_{\text{rot}}(\boldsymbol{x})$, and  $\Delta_{\text{scal}}(\boldsymbol{x})$ represent the offsets in signal strength, rotation, and scaling, respectively. Once the dynamic components are obtained, they are combined with the static components to form the complete properties of the 3D Gaussians at a specific TX position. These properties are then used in the subsequent coordinate projection transformation and electromagnetic signal calculation processes.

By incorporating deformable 3D Gaussians, the improved scenario representation network effectively integrates the characteristics of large-scale and small-scale fading into the channel model. This enhancement enables the network to accurately capture the high-frequency variations, significantly improving its interpretability and characterization capabilities.

\subsection{Electromagnetic Splatting with $\alpha$-blending}

With the introduction of a new structure in the scenario representation network, WRF-GS+ enhances the original framework by associating environmental information with the opacity attribute of 3D Gaussians. This opacity is translated into the weight of each virtual TX, which implicitly characterizes attenuation information, enabling a more accurate representation of signal propagation. Compared with WRF-GS, WRF-GS+ eliminates the need for additional attributes. This simplifies the 3D-GS model by reducing the number of parameters and memory overhead while improving convergence.

In the field of optics, opacity is modeled as particle density, which characterizes the probability of particles in space being blocked. This concept is adapted to the wireless domain, where opacity is used to represent the environmental influence on signal propagation. Specifically, the opacity of a 3D Gaussian reflects the material properties at its location, which in turn determines the weight of the corresponding virtual TX. By incorporating this property, WRF-GS+ can effectively model how environmental factors, such as obstacles or reflectors, affect signal strength at different locations. 

Based on the $\alpha$-blending technique described in Eqn. \eqref{eq:redering} for optical rendering, we extend this concept to the electromagnetic domain by adapting it to complex number calculations. This allows us to rewrite Eqn. \eqref{eq:12} and Eqn. \eqref{pixel_val} as:
\begin{equation}
    R_{k} = \displaystyle\sum_{i=1}^{N}({S}(\boldsymbol{x}_{i})+ \Delta_{\text{sig}}(\boldsymbol{x}_{i})){\alpha}_{i}\displaystyle\prod_{j=1}^{i-1}(1-{\alpha}_{j}),
\label{eq:EMredering}
\end{equation}
where $R_{k}$ represents the received signal at a specific angle (pixel) $k$.  This formula indicates that the received signal is the superposition of contributions from  Gaussians, weighted by their respective $\alpha$-blending coefficients. Each Gaussian's contribution consists of a static component ${S}(\boldsymbol{x}_{i})$ and a dynamic component $ \Delta_{\text{Sig}}(\boldsymbol{x}_{i})$, reflecting the influence of environmental and TX movement effects, respectively.

To provide an intuitive understanding of this process, we revisit the example illustrated in Fig. \ref{splatting}(c). For the pixel value $R_{1}$ in the upper-left corner, only the signal from $G(\boldsymbol{x}_{1})$ contributes, but it is now weighted by its opacity coefficient $\alpha_{1}$. For the pixel value $R_{2}$ in the upper-right corner, the contributions from $G(\boldsymbol{x}_{1})$, $G(\boldsymbol{x}_{2})$, and $G(\boldsymbol{x}_{3})$ must be considered. The coefficient for the signal from $G(\boldsymbol{x}_{1})$ remains $\alpha_{1}$, while the coefficients for $G(\boldsymbol{x}_{2})$ and $G(\boldsymbol{x}_{3})$ are adjusted to $\alpha_{2}(1-\alpha_{1})$ and $\alpha_{3}(1-\alpha_{1})(1-\alpha_{2})$, respectively. The calculation for other pixel values $R_{k}$ follows the same principle.

By leveraging the static nature of environmental information, we associate this information with the opacity attribute of 3D Gaussians, which depends only on their positions. WRF-GS+ achieves a more accurate and interpretable representation of the WRF by decomposing the virtual TX into static and dynamic components.

\subsection{Summary}
The proposed WRF-GS+ framework addresses the limitations of WRF-GS in modeling dynamic signal variations and parameter redundancy. First, leveraging insights from the different characteristics of large-scale fading and small-scale fading in wireless communication scenarios, we enhance the scenario representation network by introducing deformable 3D Gaussians. This approach can efficiently capture the high-frequency signal variations caused by complex multipath effects.
By integrating deep learning with physics-based modeling, the network achieves a more effective representation of high-frequency signal variations. Second, we enhance the electromagnetic splatting process by using the inherent opacity of 3D Gaussians as the weighting factor, eliminating the need for the network to explicitly represent attenuation parameters. This modification not only reduces computational complexity but also enhances the framework’s efficiency by leveraging the natural properties of 3D Gaussians to model signal attenuation implicitly. Finally, we implement an $\alpha$-blending method to compute the received signals from each direction in parallel, synthesizing the spatial spectrum with improved efficiency.

\section{Implementation and Evaluation}\label{SecIAE}
In this section, we implement and evaluate the WRF-GS and WRF-GS+ methods in a laboratory environment using a real-world dataset in \cite{zhao2023nerf2}.

\subsection{Implementation Details}\label{SecIAE_A0}
The proposed methods are implemented in Python, utilizing customized CUDA kernels for rasterization \cite{kerbl3Dgaussians}, and trained on an NVIDIA GeForce RTX 3090 GPU. It is worth noting that we handle complex signals by separating them into real and imaginary components via Euler's formula rather than decomposing complex signals into amplitude and phase. This operation enables parallelized computations in the CUDA kernels. The implementation details are provided below.

\textbf{Position Initialization:}
We randomly generate 3D points within a specified range. The positions of these points are used to initialize the center coordinates of the 3D Gaussians. In addition, we implement an adaptive density control strategy \cite{kerbl3Dgaussians} to learn the positional distribution of the 3D Gaussians during the training process. Compared with the LiDAR point-cloud initialization \cite{wen2025wrf}, the random point initialization can overcome the requirement for multimodal input and minimize the reliance on the quality of LiDAR point clouds.
 
\textbf{Position Encoding:}
To improve the spatial resolution, we introduce an efficient position encoding method \cite{mildenhall2021nerf}, i.e.,
\begin{equation}
\gamma (\boldsymbol{t})=(\sin(\pi \boldsymbol{t}),\cos(\pi \boldsymbol{t}),\ldots,\sin(2^{L}\pi \boldsymbol{t}),\cos(2^{L}\pi \boldsymbol{t})),
\end{equation}
where $\boldsymbol{t}$ represents the 3D coordinates input and $L$ is the order of the position encoding. For the input position encoding of $P_{\text{TX}}$ and $G(\boldsymbol{x})$ in the MLPs, we set $L = 9$.

\textbf{Optimization Details:}
In the scenario representation network, we need to train the MLPs and learn the 3D Gaussian representations using the training dataset. We adopt the Adam optimizer for training and 3D Gaussian representation. The loss function is computed by comparing the difference between the synthesized spatial spectrum $I_{\text{pred}}$ and the ground truth $I_{\text{gt}}$: 
\begin{equation}
    \mathcal{L} = (1 - \eta) |I_{\text{gt}} - I_{\text{pred}}| + \eta (1 - \xi(I_{\text{gt}}, I_{\text{pred}})),
\end{equation}
where $\xi(I_{\text{gt}}, I_{\text{pred}})$ is the structural similarity index measure (SSIM) function \cite{wang2004image}, which is used to measure the similarity between two images. In addition, $\eta = 0.2$ is a weighting factor.

\subsection{Experiment Design} \label{SecIAE_B}

\textbf{Dataset.} We use an open-source dataset provided in NeRF\textsuperscript{2} \cite{zhao2023nerf2} to evaluate the proposed frameworks. This dataset is obtained in a laboratory environment, which consists of multiple tables, shelves, and several small rooms. The 3D LiDAR point clouds of the environment are shown in Fig. \ref{SynSS}(a). The RX is equipped with a $4 \times 4$ antenna array and operated at 915MHz, while the TX is an RFID tag that sends RN16 messages repeatedly. To collect the training dataset, the position of the RX is fixed and the position of the TX is systematically varied throughout the laboratory. Each sample in the dataset consists of the TX position and the corresponding spatial spectrum measured at the RX (i.e., Eqn. \eqref{eq:ss02}).  

\textbf{Benchmarks.} We compare WRF-GS+ and WRF-GS with several baselines:
\begin{itemize}
    \item \textbf{Ray Tracing}\cite{yun2015ray}: This approach emits rays from the RX through every pixel in the laboratory using the 3D LiDAR point clouds. These rays interact with the objects, resulting in reflection, diffraction, and scattering, which are recursively traced to simulate real-world ray propagation using the RayTracing toolkit in MATLAB.
    
    \item \textbf{Variational Autoencoder (VAE)} \cite{kingma2013auto}: VAE is commonly used to generate new data samples by mapping observed data to a continuous latent space \cite{ha2020food}. Here, we use VAE to predict the spatial spectrum at arbitrary TX positions.
    
    \item \textbf{Deep Convolutional Generative Adversarial Network (DCGAN)} \cite{radford2015unsupervised}: DCGAN comprises a generator and a discriminator that are trained in an adversarial manner. The generator learns to generate realistic samples to deceive the discriminator, while the discriminator aims to distinguish the real samples from the generated ones. 
    
    \item\textbf{NeRF\textsuperscript{2}} \cite{zhao2023nerf2}: NeRF\textsuperscript{2} is a method for spatial spectrum synthesis using the NeRF technique. Similarly to our work, NeRF\textsuperscript{2} can synthesize the spatial spectrum at arbitrary TX positions after training in a given scenario.
\end{itemize}

\subsection{Evaluation}
We first compare the accuracy of the spatial spectrum synthesis generated by the WRF-GS+, WRF-GS, NeRF\textsuperscript{2}, VAE, and ray tracing methods, as shown in Fig. \ref{SynSS}(b). The visualization is obtained by transforming the matrix form in Eqn. \eqref{eq:ss02} into a polar coordinate representation.  From Fig. \ref{SynSS}(b) we see that WRF-GS+ achieves the best performance compared with the WRF-GS method and the baselines, closely matching the ground truth.

\begin{figure}[!t]
    \centering
    \includegraphics[width=0.48\textwidth]{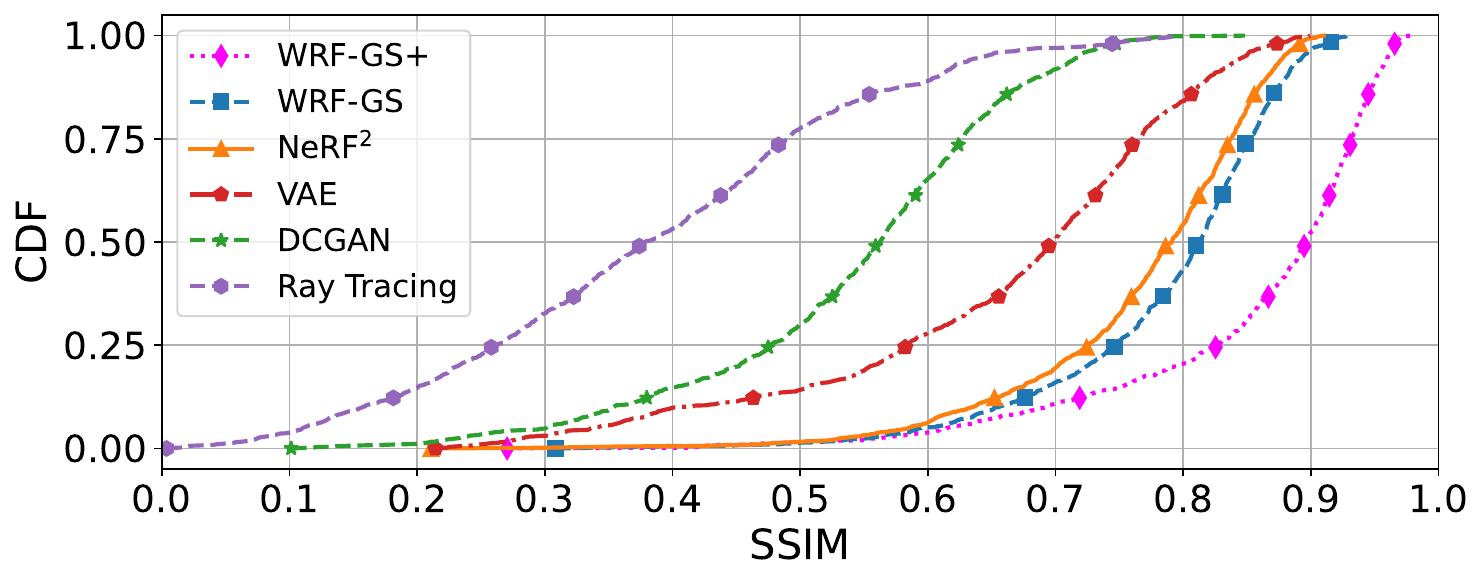}
    \caption{The SSIM values of the synthesized spatial spectra generated by the WRF-GS+, WRF-GS, NeRF\textsuperscript{2}, VAE, DCGAN, and ray tracing methods.}
    \label{ssim}
\end{figure}

Next, we compare the SSIMs of the synthesized spatial spectra generated by different methods. The value of SSIM ranges from $0$ to $1$, where a higher value indicates greater similarity between the synthesized and ground truth samples. The cumulative distribution function (CDF) of SSIM illustrates the disparities between the methods and the ground truth. From Fig. \ref{ssim}, we see that the median SSIM values for WRF-GS+, WRF-GS, NeRF\textsuperscript{2}, VAE, DCGAN, and ray tracing are $0.90$, $0.82$, $0.78$, $0.70$, $0.56$, and $0.38$, respectively, and their 90th percentiles are $0.95$, $0.88$, $0.86$, $0.82$, $0.69$, and $0.61$. Evidently, WRF-GS+ significantly outperforms WRF-GS thanks to its adaptation of deformable 3D Gaussians in the scenario representation network design. This operation enables the WRF-GS+ method to efficiently characterize the distribution of the WRF in a given scene by separately modeling the static and dynamic components of the signals from virtual TXs. In addition, the poor performance of NeRF\textsuperscript{2} is due to its reliance on the voxel sampling rate along the pixel ray. As a result, a high resolution results in significant computational overhead and demands extensive training data, while a lower resolution compromises the representation of WRF. Thus, there is a trade-off between representation capability and computational efficiency. In contrast, the ray tracing method achieves the poorest performance as it heavily relies on environmental material information, which is often hard to obtain.

\begin{figure}[!t]
    \centering
    \includegraphics[width=0.48\textwidth]{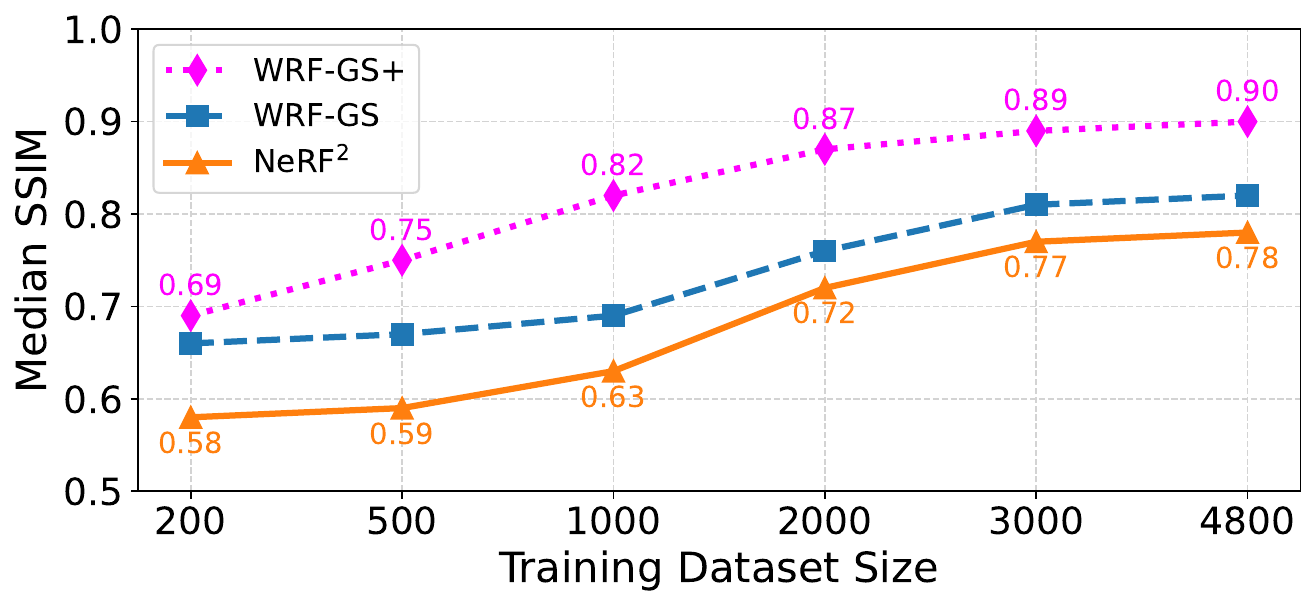}
    \caption{The median SSIM values of the WRF-GS+, WRF-GS and NeRF\textsuperscript{2} methods under different training dataset sizes.}
    \label{datasize}
\end{figure}
Fig. \ref{datasize} compares the median SSIM values of the WRF-GS+, WRF-GS, and NeRF\textsuperscript{2} methods across varying sizes of the training dataset. We maintain a consistent test dataset and randomly subsample the original training dataset to create new training datasets ranging from $200$ to $4,800$ samples. It can be seen that  WRF-GS+ consistently outperforms the WRF-GS and NeRF\textsuperscript{2} methods for all training dataset sizes, improving the median SSIM by $3\%$ to $13\%$ and $11\%$ to $19\%$, respectively. This demonstrates that WRF-GS+ has the highest sample efficiency. More importantly, the performance gap increases as the training dataset size decreases, which indicates that the proposed methods require less training data and, consequently, less time and effort from experienced engineers.

\begin{figure}[!t]
\centering
\includegraphics[width=0.48\textwidth]{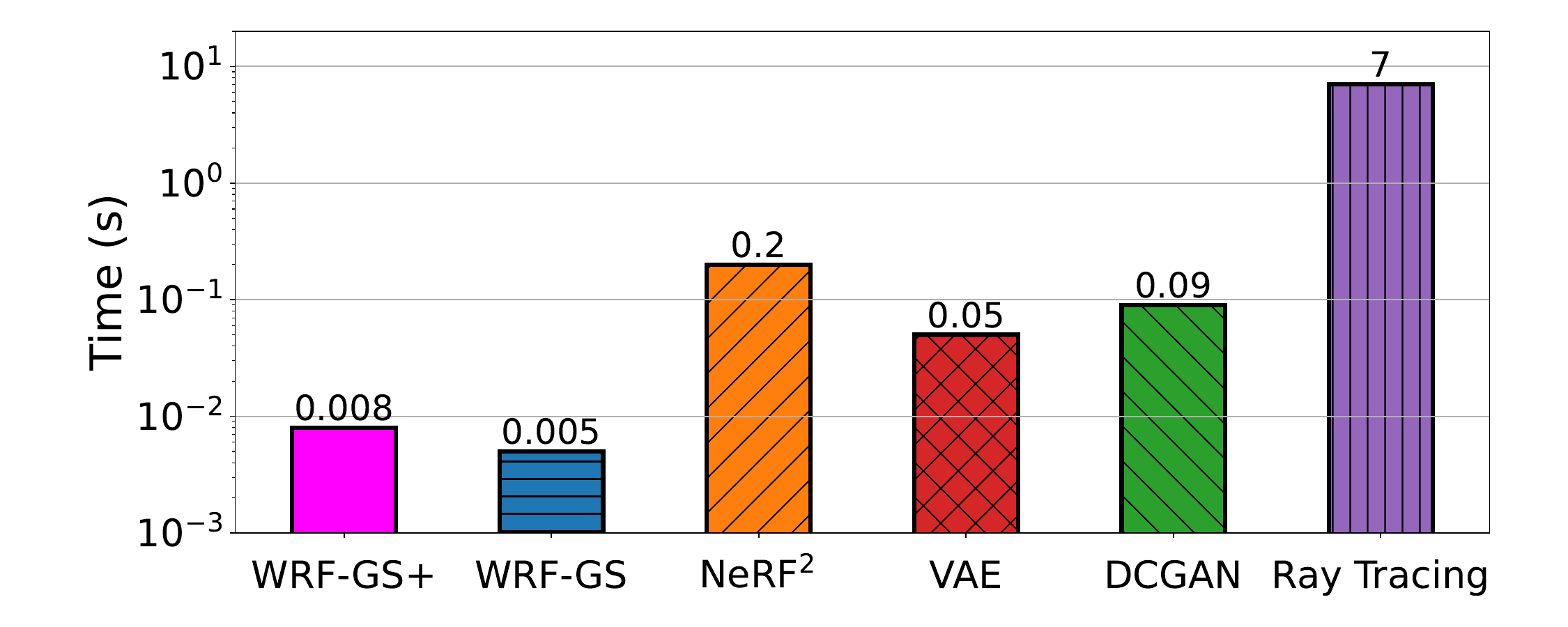}
    \caption{The rendering time required to synthesize a spatial spectrum using different methods.}
    \label{speed}
\end{figure}
Fig. \ref{speed} illustrates the rendering time required to synthesize a spatial spectrum using different methods under the same settings. The results show that WRF-GS+ achieves a processing time of $0.008$s per sample, while WRF-GS, NeRF\textsuperscript{2}, VAE, DCGAN, and Ray Tracing require $0.005$s, $0.2$s, $0.05$s, $0.09$s, and $7$s, respectively. Notably, the processing times of WRF-GS+ and WRF-GS are one to three orders of magnitude lower than those of the other methods. This significant advantage makes the proposed methods highly suitable for latency-sensitive applications. Although the processing time of WRF-GS+ is slightly higher than that of WRF-GS, this difference is primarily due to the inclusion of the deformation network in WRF-GS+, which introduces more parameters compared to the two MLPs used in WRF-GS. However, given the significant improvement in spectrum synthesis accuracy achieved by WRF-GS+, this millisecond-level difference is negligible and acceptable in most practical application scenarios.

\section{Case Studies}\label{SecCS}
In this section, we further demonstrate the effectiveness of the proposed methods on RSSI and CSI prediction tasks.

\subsection{Case Study I: RSSI Prediction}\label{SecRSSIP}
We first apply the WRF-GS and WRF-GS+ methods to the RSSI prediction task, which involves estimating the signal strength at a specific location based on factors such as distance, environmental conditions, and prior measurements. RSSI prediction is essential for network management, indoor positioning, and signal coverage planning. For example, the RSSI prediction accuracy directly affects the positioning accuracy of the localization task and the performance of the base station (BS) deployment in cellular networks.

Unlike the previous spatial spectrum reconstruction task, the RSSI measurements are obtained by a single omnidirectional antenna instead of an antenna array. As a result, the output is a composed signal containing the power from all directions rather than the $90\times 360$ pixel matrix. To complete the RSSI prediction task, the proposed methods are trained on the public BLE dataset \cite{zhao2023nerf2}. We compare the proposed methods with the NeRF\textsuperscript{2} method and the MRI \cite{shin2014mri} method. MRI is a learning-based method that interpolates the RSSI values at the unsampled location using a log-distance path-loss model.

\subsubsection{Experiment Setup}
The BLE dataset is obtained in a scene containing several rooms with a total area of about $15,000 ft^{2}$. There are $21$ RXs and a TX moving in this area. The RSSI at these RXs is recorded when the TX moves from one position to another. There are a total of $6,000$ RSSI measurements in the BLE dataset. Note that we consider the measurement to be an invalid value when the RX is far from the TX. In this situation, the recorded RSSI is set to $-100$ dBm. There are $2,000$ valid RSSI measurements for each RX, where $1,600$ measurements are used for training and the remaining $400$ are used for testing.

\subsubsection{RSSI Prediction Accuracy}
\begin{figure}[!t]
    \centering
    \includegraphics[width=0.48\textwidth]{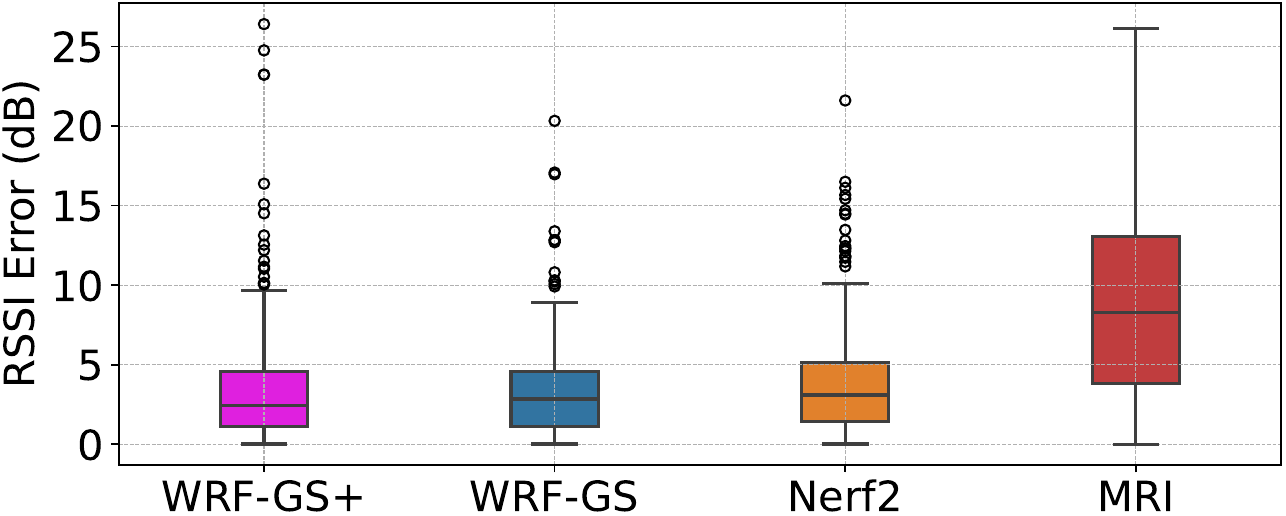}
    \caption{The RSSI prediction errors of the WRF-GS+, WRF-GS, NeRF\textsuperscript{2}, and MRI methods on the public BLE dataset.}
    \label{RSSI}
\end{figure}
Fig. \ref{RSSI} compares the RSSI prediction errors of the WRF-GS+, WRF-GS, NeRF\textsuperscript{2}, and MRI methods on the public BLE dataset. The prediction error is defined as the absolute difference between the predicted and ground truth RSSI values. In simulation, we train the proposed methods on the dataset of each RX. The result measured in dB is the average value of the RSSI prediction errors for $12$ RXs. From Fig. \ref{RSSI}, we see that WRF-GS+ has the lowest RSSI prediction error among the WRF-GS, NeRF\textsuperscript{2}, and MRI methods.
The median of WRF-GS+ is $2.4$ dB ($10$-th percentile: $1.1$ dB; $90$-th percentile: $4.6$ dB). In contrast, the median errors of WRF-GS, NeRF\textsuperscript{2}, and MRI are $2.9$ dB, $3.1$ dB, and $8.3$ dB, respectively. The proposed methods clearly outperform the baselines, indicating that the 3D-GS-based method is more suitable than the NeRF-based method in this task. The superior performance of the WRF-GS+ is due to its ability to separate the signals from virtual TXs into static and dynamic components. This operation enables an accurate RSSI prediction even with continuous TX movements.

\subsection{Case Study II: CSI Prediction}\label{SecCSIP}
We adapt the WRF-GS+ and WRF-GS methods to the downlink CSI prediction task in MIMO systems. Typically, the downlink and uplink transmission are operated on different frequency bands in frequency domain duplex systems, where the uplink-downlink reciprocity does not hold. The BS can obtain the downlink CSI of each antenna from the uplink channel feedback by sending pilot sequences to the end devices. However, the overhead of such feedback is linearly proportional to the number of antennas and devices, making it impossible in massive MIMO systems. 

Motivated by the fact that the uplink and downlink transmissions experience the same physical environment, it is reasonable to use the uplink CSI measured at the BS to infer the downlink CSI. To achieve this goal, we employ WRF-GS to predict the downlink CSI based on the uplink CSI. As pointed out in \cite{xie2019md}, each CSI is unique and highly correlated with the physical environment. Therefore, there is a mapping between the device's location and its uplink CSI, which is similar to fingerprint-based localization. Accounting for this, Eqn. \eqref{F_theta} can be rewritten as
\begin{equation}
    {F}_{\Theta}:(G(\boldsymbol{x}),I_{u}(\boldsymbol{x}))\Rightarrow\left(\delta(\boldsymbol{x}), S(\boldsymbol{x})\right),
\end{equation}
where $I_{u}(\boldsymbol{x})$ is the uplink CSI at location $\boldsymbol{x}$. Thus, we can apply the WRF-GS framework to the downlink CSI prediction task by employing a similar configuration in Section \ref{SecIAE_A0}. 

For the WRF-GS+ method, we adapt the deformation network to this CSI prediction task by modifying Eqn. \eqref{D_theta} into
\begin{equation}
    {D}_{\Theta}:(G(\boldsymbol{x}), I_{u}(\boldsymbol{x}))\Rightarrow\left(  \delta_{\text{Sig}}(\boldsymbol{x}), \delta_{\text{r}}(\boldsymbol{x}),  \delta_{\text{s}}(\boldsymbol{x})   \right).
\label{D_theta_csi}
\end{equation}
Note that the input is changed from the TX's location to the uplink CSI, and the output of the whole pipeline is changed from the spatial spectrum to the downlink CSI.

\begin{figure}[!t]
    \centering    

    \includegraphics[width=0.48\textwidth]{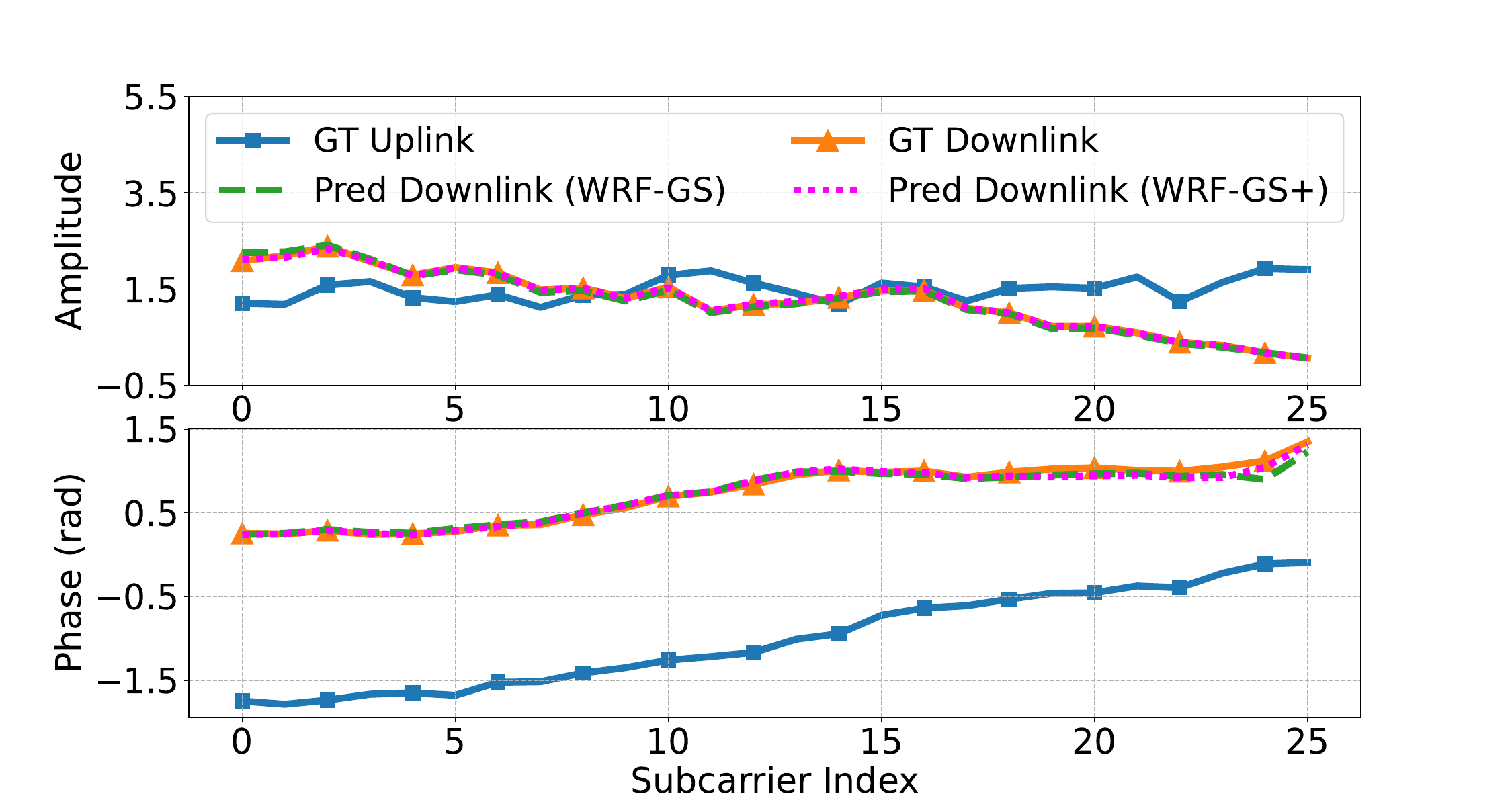}
    \caption{Comparison of the ground truth CSI and the predicted CSI using the proposed methods in terms of amplitude and phase via different sub-carriers.}
    \label{Fig:CSI}
\end{figure}

\subsubsection{Experiment Setup}
We train the WRF-GS and WRF-GS+ methods on the public Argos channel dataset \cite{shepard2016understanding} for the downlink CSI prediction task. This dataset is collected in a real environment. The CSI is measured in different environments where a BS is equipped with $104$ antennas and serves multiple users. Each CSI measurement contains $52$ subcarriers. Similarly to \cite{zhao2023nerf2} and \cite{2021Liu}, we regard the first $26$ subcarriers as the uplink channel and the remaining $26$ subcarriers as the downlink channel.
Unlike the spatial spectrum reconstruction task, which requires computing power from all directions separately, we focus on a complex value of the predicted CSI at each subcarrier. Notice that this complex value is a composed signal that sums in all directions. Apart from NeRF\textsuperscript{2}, we also compare the proposed methods with the following baselines:

\begin{itemize}
    \item \textbf{FIRE} \cite{2021Liu}: FIRE is a deep learning method based on the VAE architecture and is used to predict the downlink CSI by learning the latent distribution of the uplink CSI.
    \item \textbf{R2F2} \cite{vasisht2016eliminating}: Given the uplink CSI, R2F2 can obtain each path's parameters and the number of paths by solving an optimization problem to estimate the downlink CSI.
    \item \textbf{OptML} \cite{bakshi2019fast}: OptML is a deep learning method based on the R2F2 framework and can be used to predict the underlying path information of the downlink channel.
\end{itemize}

\subsubsection{Channel Estimation Accuracy}
Fig. \ref{Fig:CSI} depicts the prediction accuracy of the WRF-GS and WRF-GS+ methods in the downlink CSI prediction task. Since the predicted CSI is a complex value, we measure the prediction accuracy on the amplitude and phase parts. We see that the predicted results are close to the ground truth. This means that the proposed methods can successfully predict the downlink CSI by extracting information from the uplink CSI based on the neural network architecture and 3D-GS.

\begin{figure}[t]
    \centering   
    \includegraphics[width=0.48\textwidth]{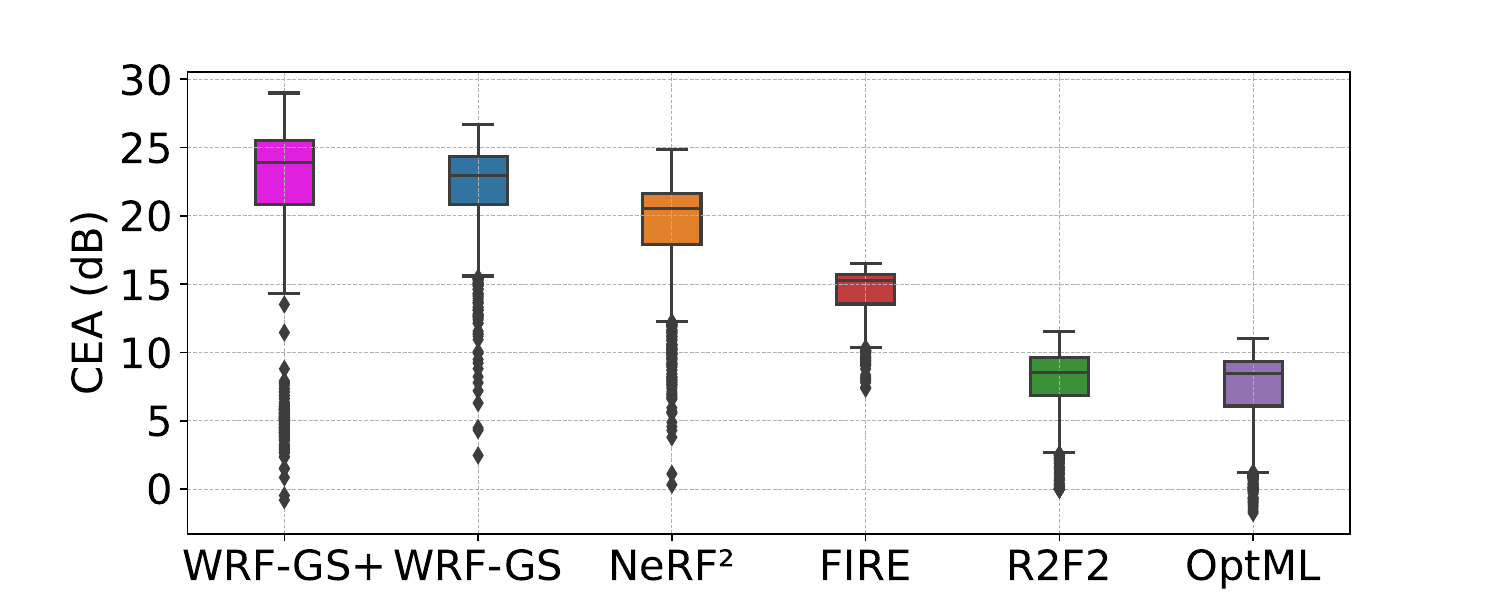}
    \caption{The prediction accuracy of different methods in the downlink CSI prediction task. The figure shows the maximum, minimum, median, upper and lower quartile ranges of the CEA distribution for different methods, as well as the outliers. The higher the CEA, the better the prediction accuracy.}
    \label{Fig:snr}
\end{figure}

Next, we adopt the following metric to quantify the prediction accuracy and refer to it as CEA, representing the channel estimation accuracy in dB:
\begin{equation}
   \text{CEA} = -10 \log _{10}\left( \frac{\left\|\mathrm{} C_{\text{pred}}-C_{\text{GT}}\right\|^{2}}{\left\|C_{\text{GT}}\right\|^{2}}\right),
\end{equation}
where $C_{\text{pred}}$ is predicted downlink channel and $C_{\text{GT}}$ is ground truth downlink channel. The higher the CEA, the better the prediction accuracy. Fig. \ref{Fig:snr} compares the prediction accuracy of the proposed methods and baselines. The results show that WRF-GS+ achieves a median CEA of $23.91$ dB ($10$th percentile: $20.85$ dB, $90$th percentile: $25.52$ dB), while WRF-GS achieves a median CAE of $22.98$ dB ($10$th percentile: $20.82$ dB, $90$th percentile: $24.33$ dB). In contrast, the median CEAs of NeRF\textsuperscript{2}, FIRE, R2F2, and OptML are $20.55$ dB, $15.29$ dB, $8.57$ dB, and $8.47$ dB, respectively. The proposed methods consistently outperform the baselines. Notably, WRF-GS+ achieves the highest prediction accuracy, surpassing the best baseline NeRF\textsuperscript{2} by a significant margin of $3.36$ dB.
This demonstrates that the superior capability of the proposed methods in efficiently and accurately modeling wireless channels.

\section{Conclusion}\label{SecCON}
This paper first introduced WRF-GS, a novel framework for wireless channel modeling based on WRF reconstruction using neural network and 3D-GS. WRF-GS effectively captures the interactions between the environment and radio signals by refining the optical 3D-GS technique from three key aspects: scenario representation network, projection model, and electromagnetic splatting. To improve efficiency and accuracy, we further proposed WRF-GS+ by incorporating deformable 3D Gaussians and $\alpha$-blending to the scenario representation network and electromagnetic splatting modules, respectively. Experimental results showed that the proposed methods outperform the baselines for spatial spectrum synthesis. They can synthesize new channel characteristics within milliseconds using a small number of training samples. Moreover, WRF-GS+ achieves superior performance in both RSSI and CSI prediction tasks, surpassing existing methods by more than $0.7$ dB and $3.36$ dB, respectively. We envision that the proposed methods will enable various applications in future networks by leveraging their high-fidelity visualization, low sample complexity, and rapid rendering speed.

\bibliographystyle{./IEEEtran}
\bibliography{./IEEEabrv,./3DGS}
\end{document}